\documentclass[aps,prb,twocolumn,amsmath,amssymb,superscriptaddress]{revtex4}

\usepackage{amssymb}
\usepackage{amsmath}
\usepackage{graphicx}
\usepackage{textcomp}
\usepackage{color}
\usepackage{amsfonts}
\usepackage{epsfig}
\usepackage{hyperref}
\usepackage{bm}
\usepackage[caption=false]{subfig} 
\usepackage{booktabs}
\usepackage{tabularx}
\usepackage{array}
\usepackage{soul}
\usepackage{parskip}

\newcommand{\arctanh}[1]{\operatorname{arctan}}

\newcolumntype{M}[1]{>{\centering\arraybackslash}m{#1}}

\DeclareMathAlphabet\mathbfcal{OMS}{cmsy}{b}{n}

\begin{document}

\title{Dynamical Mean-Field Theory for spin-dependent electron transport in spin-valve devices}

\author{Andrea Droghetti}
\email{andrea.droghetti@tcd.ie}
\affiliation{School of Physics and CRANN, Trinity College, Dublin 2, Ireland}
\author{Milo\v s M. Radonji\'c}%\footnote[3]{These authors contributed equally to this work.}}
\affiliation{Institute of Physics Belgrade, University of Belgrade, Pregrevica 118, 11080 Belgrade, Serbia}
\author{Liviu Chioncel}
\affiliation{Theoretical Physics III, Center for Electronic Correlations and Magnetism,
Institute of Physics, University of Augsburg, 86135 Augsburg, Germany}
\author{Ivan Rungger}
\affiliation{National Physical Laboratory, Hampton Road, Teddington TW11 0LW, United Kingdom}

\begin{abstract}
We present the combination of Density Functional Theory (DFT) and Dynamical Mean Field Theory (DMFT) for computing the electron transmission through two-terminals nanoscale devices. The method is then applied to metallic junctions presenting alternating Cu and Co layers, which exhibit spin-dependent charge transport and giant magnetoresistance (GMR) effect. 
 The calculations show that the coherent transmission through the $3d$ states is greatly suppressed by electron correlations. This is mainly due to the finite lifetime induced by the electron-electron interaction and is directly related to the imaginary part of the computed many-body DMFT self-energy. At the Fermi energy, where in accordance with the Fermi-liquid behaviour the imaginary part of the self-energy vanishes, the suppression of the transmission is entirely due to the shifts of the energy spectrum induced by electron correlations. Based our results, we finally suggest that the GMR measured in Cu/Co heterostructures for electrons with energies about 1 eV above the Fermi energy is a clear manifestation of dynamical correlation effects.
\end{abstract}

\maketitle
\section{Introduction}
Spintronics\cite{Zu.Fa.04} employs the electron spin for sensing and information technology applications.
%and emerged from the study of spin-dependent electron transport phenomena in solid-state systems. 
The prototypical spintronic device is the spin-valve. It consists of two or more conducting ferromagnetic layers - typically $3d$ transition metals (TMs), 
whose electrical resistance changes depending the relative alignment of the layers' magnetization\cite{Gr.08,Fe.08}. 
This phenomenon is called giant magnetoresistance (GMR) effect and is  exploited in the read heads of hard-disk drives.
%The origin of the GMR is the dependence of the electrical conduction in ferromagnetic materials on the spin state of the electrons.
%The experimental results are generally accounted for in terms of the phenomelogical “two current model”, 
%which assumes that majority and minority electrons have different conductivities
%and the different spin channel do not mix. 
The GMR is due to the different conductance of the majority and of the minority electrons in ferromagnets.
The earlier GMR experiments\cite{bi.gr.89,ba.br.88} were conducted with the so-called ``current-in-plane'' configuration, whereas recent experiments use thin film heterostructures, where the current flows perpendicular to the various layers' planes, achieving higher performances\cite{ba.pr.99}.\\
Over the last two decades there has been considerable progress in the computational modeling of current-perpendicular-to-plane spin-valves. 
In particular, the ballistic transport properties have been addressed by using the Landauer-B\"uttiker formalism\cite{La.57,Bu.86,Bu.88}, where
the conductance is determined by the electron transmission through the device region placed between two semi-infinite electrodes.
The transmission is calculated via the tight-binding approach\cite{Ts.Pe.97,Sa.La.99}, or, in first-principles studies, via Kohn-Sham density functional theory (DFT)\cite{jo.gu.89,kohn.99,jone.15} within the local spin density approximation (LSDA)\cite{ba.he.72,vo.wi.80} or the generalized gradient approximation (GGA)\cite{pe.ch.92,pe.ch.93,pe.bu.96}. Various
implementations exist, based on transfer matrix \cite{Wo.Is.02,Wo.Is.02_2, ku.dr.00}, layer-Korringa–Kohn–Rostoker (KKR) \cite{Ma.Zh.99}, mode-matching\cite{Kh.Br.05} or non-equilibrium Green’s function (NEGF) techniques\cite{Ta.Gu.01,Ba.Mo.02,ro.ga.06, book1}. The main assumption is that all
the materials in a device can be treated at the  effective single-particle level, and that the DFT Kohn-Sham band structure provides an accurate first-principles representation of quasi-particle spectral properties.
%The main assumption beyond the use of DFT for first-principles transport calculations is that the effective Kohn-Sham single-particle band structure provides an accurate representation of the quasi-particle spectral properties of a material. However, this does not hold true for ferromagnetic $3d$ TMs used in spin-valves since they are moderately correlated \cite{ka.li.02}. The KS DFT band structure drastically overestimates the spin splitting of the $3d$ bands\cite{mo.ma.02,br.mi.06}, it presents too wide majority spin bands, it does not account for the intrinsic broadening of quantum states, and it does not display spectroscopic features, such as satellites~\cite{gu.ba.77}.
However, this may not hold true for ferromagnetic $3d$ TMs used in spin-valves since they are moderately correlated \cite{ka.li.02}. %The KS DFT band structure drastically overestimates the spin splitting of the $3d$ bands\cite{mo.ma.02,br.mi.06}, it presents too wide majority spin bands, it does not account for the intrinsic broadening of quantum states, and it does not display spectroscopic features, such as satellites~\cite{gu.ba.77}. 
To our knowledge, no first-principles studies have addressed the impact of electron correlations on the transport properties of spin-valves. 
In light of this, the goal of our work is to present a computational platform to compute the transmission coefficient and the GMR of two-terminal spintronic devices with electronic spectra treated beyond the single-particle Kohn-Sham DFT picture. \\
A significant progress in the theoretical understanding of correlation effects in materials has been achieved with the dynamical mean-field theory (DMFT)~\cite{me.vo.89,ge.ko.96,ko.vo.04,ko.sa.06}. 
In the so-called LSDA+DMFT scheme~\cite{ko.sa.06,held.07}, LSDA calculations provide the material dependent inputs (orbitals and hopping parameters) from first-principles, while DMFT solves the many-body problem for the local interactions. In the case of $3d$ ferromagnetic TMs,
LSDA+DMFT has been applied to address spectral properties of bulk materials~\cite{li.ka.01,gr.ma.07} and surfaces\cite{gr.ma.07}, digital  heterostructures~\cite{be.ho.11,ch.le.11}, alloys~\cite{os.vi.18}, interfaces containing half-metallic ferromagnets~\cite{ma.he.18,ke.ma.20} and to estimate magnetic moments above and below the Curie temperature~\cite{li.ka.01}. 
 In all these studies LSDA+DMFT provides qualitative and quantitative improvements over DFT for the description of the systems electronic and magnetic properties. 
 In the context of two-terminal devices, LSDA+DMFT has been applied to compute the linear-response conductance of point contacts\cite{ja.ha.09,ja.ha.10,Ja.15}, molecular junctions\cite{ja.so.13,dr.ru.17,ap.dr.18,ja.18,ru.ba.19,bh.to.21}, and heterostructures comprising a single correlated layer\cite{ch.mo.15,mo.ap.17}, but, to our knowledge, never to address spin-dependent effects in TM-based spin-valves.\\
In this paper, we describe the integration of the LSDA+DMFT framework within the Smeagol quantum transport code\cite{ro.ga.06,ru.sa.08}. In particular, we generalize layer-DMFT \cite{ok.mi.04,po.no.99,va.am.18,we.ap.21} towards first-principles calculations for perpendicular-to-the-plane spintronic devices in the, so-called, zero-bias limit. Our implementation is based on the NEGF method to obtain the spin-dependent transmission coefficient through a correlated region attached to two semi-infinite electrodes.
Second order perturbation theory in the screened electron-electron interaction $U$ is employed as DMFT solver, allowing for the fast evaluation of the self-energy directly on the real frequency axis with no need of analytic continuation schemes. LSDA+DMFT transport calculations carried out by means our implementation are, in practice, only slightly more computationally demanding than standard DFT+NEGF calculations. The solver is accurate for moderately correlated materials such as ferromagnetic TMs, where $U$ is smaller than the band-width. Nonetheless, our implementation can be easily extended to include any other solver, and therefore also allowing one to treat strongly correlated systems. We expect that this will pave the way towards systematic studies of correlation effects in quantum transport. \\
The performance of our method is illustrated in detail for a number of prototypical spintronic heterostructures with alternating Cu and Co layers, where electrons are correlated in the $3d$ orbitals. We demonstrate that electron correlations drastically affect the transmission coefficient, the energy level alignment between the $3d$ and $s$ states, and, therefore, zero-bias transport properties. Moreover, we suggest that the GMR effect, which has been measured in hot electron transport experiments\cite{ka.ro.08}, is a striking manifestation of electron correlation.  \\
The paper is organized as follows. To begin with, in Sec. \ref{sec.DFTNEGF}, we review the basic theory of quantum transport formulated in terms of the NEGF, and its combination with DFT. Then, in Sec. \ref{Sec.Transport.Corr}, we extend the NEGF technique to systems, for which an effective single-particle picture is not appropriate. We present our numerical implementation of DMFT in Secs. \ref{Sec.Projection1}, \ref{Sec.Projection2} and \ref{Sec.DMFT}, and the basic equations for the perturbative impurity solver in Sec. \ref{Sec.impurity}. The computational details are given in Sec. \ref{Sec.Details}. The results are presented in Sec. \ref{Sec.Results}. Specifically, we describe the DFT and LSDA+DMFT calculations for a single Co layer sandwiched between two Cu electrodes in Sec. \ref{Sec.monolayer}, and we study the GMR effect in a complex heterostructure in Sec. \ref{Sec.spin_valve}. Finally we present our conclusions. 
 
\begin{figure}[h]
\centering\includegraphics[width=0.4\textwidth,clip=true]{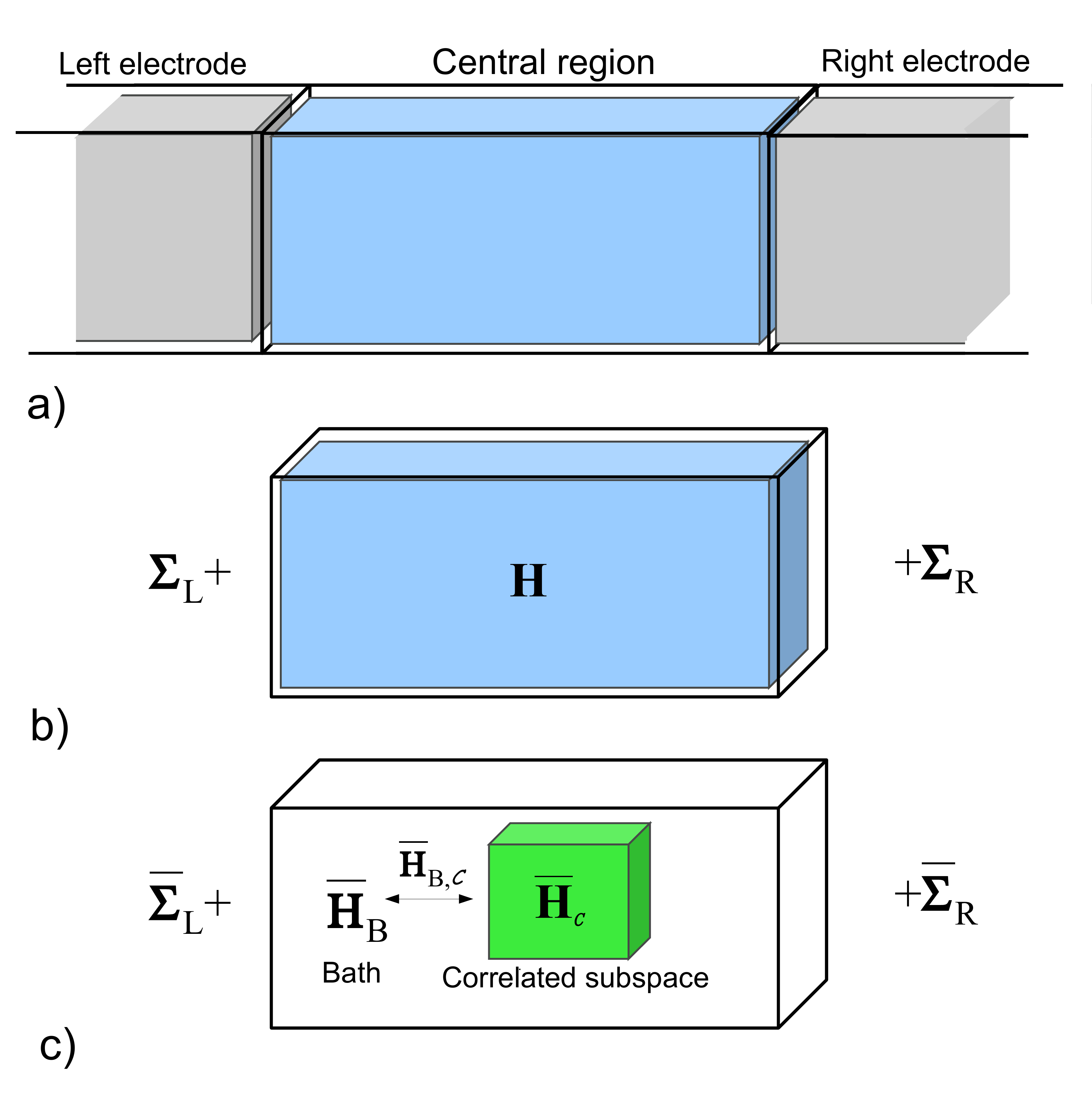}
\caption{(a) Schematic representation of a two-terminal device, which includes a central region (CR) placed between two semi-infinite electrodes. (b) The CR has Hamiltonian $H^\sigma(\mathbf{k})$, and the effect of the electrodes on the central region is captured via the left and right electrode self-energies. Note that we do not indicate the spin index and the $\mathbf{k}$-dependence in the picture to maintain the notation lighter. (c) The correlated subspace of Hamiltonian $\bar{H}^\sigma_{\mathcal{C}}(\mathbf k)$ can be separated from the rest of the CR, which we refer to as the bath. The correlated subspace and the bath are coupled through the coupling Hamiltonian $\bar{H}^\sigma_{B,\mathcal{C}}(\mathbf k)$.}
\label{pic.Device}
\end{figure}

\section{Method and implementation}
\subsection{Transport via DFT+NEGF}\label{sec.DFTNEGF}
The typical system that we consider is shown schematically in Fig. \ref{pic.Device} and represents a two-terminal device. The transport direction is along the $z$ Cartesian axis.
We employ a linear combination of atomic orbitals (LCAO) basis set.
The system is divided in three parts: a central region (CR) 
and left ($L$) and right ($R$) electrodes, from which electrons flow in and out. 
To start with, we assume that electrons in both the CR and the electrodes are effectively non-interacting.   
Each electrode is semi-infinite and periodic away from the CR; 
$\mathbf{k}=(k_x,k_y)$ indicates the wave-vector in the transverse direction.
$H^\sigma(\mathbf{k})$ is the $\mathbf{k}$-dependent single-particle Hamiltonian of the CR for electrons of spin $\sigma=\uparrow, \downarrow$. Note that we assume that there are no spin-mixing terms in the Hamiltonian. Since in general the basis set is non-orthogonal, there is also a spin-independent orbital overlap $S(\mathbf{k})$ of the CR.
We denote with $N$ the number of basis orbitals of the CR. $H^\sigma(\mathbf{k})$ and $S(\mathbf{k})$ are therefore matrices of dimension $N\times N$.
Each electrode is in local thermal
equilibrium at its own chemical potential $\mu_{L/R}$ due to its infinitely large size. When there is no applied bias voltage across the electrodes, we have $\mu_{L}=\mu_R=E_F$, where $E_F$ the Fermi energy. When a finite bias voltage $V$ is applied, the chemical potentials are shifted as $\mu_{L/R}=E_F\pm eV/2$, where $e$ is the electron charge.\\
To describe the electronic structure and the quantum transport properties of the device, we use the non-equilibrium Green's function (NEGF) approach\cite{Datta}.
The effect of electrodes on the CR is then captured via the momentum- and energy-dependent retarded electrode self-energies, $\Sigma^\sigma_L(\mathbf{k},E)$ and $\Sigma^\sigma_R(\mathbf{k},E)$. Their antihermitian parts 
\begin{equation}
\Gamma^\sigma_{L/R}(\mathbf{k},E)=i[\Sigma^\sigma_{L/R}(\mathbf{k},E)-\Sigma^\sigma_{L/R}(\mathbf{k},E)^\dagger]
\end{equation}
represent the strength of the electronic coupling between the electrodes and the CR.
%(note that, in principles, the electrodes self-energies depend also on the voltage bias\cite{book1}, but we here neglect that).
The properties of the CR are then fully described by the retarded and lesser CR Green's functions, defined as
\begin{equation}
 g^\sigma(\mathbf{k}, E)= [ES(\mathbf{k})-H^\sigma(\mathbf{k}) -\Sigma^\sigma_L(\mathbf{k},E)-\Sigma^\sigma_R(\mathbf{k},E)]^{-1},\label{retardedGF}
\end{equation}
 \begin{equation}
 g^{\sigma <}(\mathbf{k}, E)=g^\sigma(\mathbf{k}, E)[\Sigma^{\sigma<}_L(\mathbf{k},E)+\Sigma^{\sigma<}_R(\mathbf{k},E)]g^\sigma(\mathbf{k}, E)^\dagger,
\label{lesserGF}
 \end{equation} 
where the lesser electrodes' self-energies are
\begin{equation}
 \Sigma^{\sigma<}_{L(R)}(\mathbf{k},E)=if_{L(R)}(E)\Gamma^\sigma_{L(R)}(\mathbf{k},E),\label{eq.Sigma_less}
\end{equation}
with $f_{L/R}(E)=[e^{\beta(E-\mu_{L/R})}+1]^{-1}$ the Fermi function of the left/right electrode; here $\beta=1/k_\mathrm{B} \theta$, with $k_\mathrm{B}$ representing the Boltzmann constant, and $\theta$ the electronic temperature. 
The Green's functions of the CR and the self-energies are $N\times N$ matrices like $H^\sigma(\mathbf{k})$.\\
The density matrix of the CR is given by
\begin{equation}
 \rho^\sigma=\frac{1}{N_\mathbf{k}}\sum_{\mathbf{k}}\bigg[\frac{1}{2\pi i}\int dE\; g^{\sigma <}(\mathbf{k}, E)\bigg] \label{eq.DFT_rho}
\end{equation}
where $N_\mathbf{k}$ is the number of $\mathbf{k}$-points in the Brillouin zone. In our calculations, $H^\sigma(\mathbf{k})$ is the DFT Kohn-Sham Hamiltonian within the LSDA, and it is therefore density dependent. 
Eqs. (\ref{retardedGF}), (\ref{lesserGF}) and (\ref{eq.DFT_rho}) need to be evaluated self-consistently \cite{Ba.Mo.02,ro.ga.06,book1}. 
This method is usually called DFT+NEGF, and it is the state-of-the-art approach to study spin-dependent transport through nano-devices. 
Here, we use the implementation of DFT+NEGF in the Smeagol transport code\cite{ro.ga.06,ru.sa.08}, which obtains the LSDA Kohn-Sham Hamiltonian from the DFT package Siesta\cite{so.ar.02}.\\ 
The current across the CR for electrons of spin $\sigma$ is evaluated as\cite{Datta,Ba.Mo.02,ro.ga.06,book1}
\begin{equation}
 I=  \frac{e}{h}\sum_\sigma\int T^\sigma(E)[f_{L}(E)-f_{R}(E)]dE,\label{I_coherent}
\end{equation}
where $h$ is Planck's constant, and
\begin{eqnarray}
\label{TRC}
& T^\sigma(E)=\frac{1}{N_{\mathbf{k}}} \sum_\mathbf{k} T^\sigma(\mathbf{k},E)\label{eq.TRC},\\
\notag &  T^\sigma(\mathbf{k},E)= \mathrm{Tr}\Big[\Gamma^\sigma_{L}(\mathbf{k},E)  g^\sigma(\mathbf{k}, E)^\dagger  \Gamma^\sigma_{R}(\mathbf{k},E) g^\sigma(\mathbf{k}, E)\Big]
  \end{eqnarray}
is the spin- and energy-dependent transmission coefficient. The device conductance is finally defined as $ \mathfrak{G}=dI/dV$.\\ 
The approach described so far is general and can be applied to any electronic system. However, in the rest of this paper, our focus will be on metallic heterostructures. Because of electronic screening, there can be no bias voltage drop across the electrodes. Therefore, we take the linear-response limit, also denoted as zero-bias limit, $\mu_L-\mu_R\rightarrow 0$. At zero temperature ($\theta=0$)
the conductance reduces to the well-known Landauer formula
\begin{equation}
 \mathfrak{G} = \frac{\mathfrak{G}_0}{2}\sum_\sigma T^\sigma(E_F),\label{conductance}
\end{equation}
where $\mathfrak{G}_0=2e^2/h$ is the quantum of conductance. 
Notably, for non-interacting electrons, the Landauer and the so-called Kubo approaches
are equivalent\cite{ba.st.89,fi.le.81}, so that the linear-response transport properties of a system can be computed by either approach.

\subsection{Transport beyond the single-particle picture}\label{Sec.Transport.Corr}
The described NEGF method for transport is formally extended beyond the effective single-particle picture by adding the many-body retarded and lesser self-energies,
$\Sigma^\sigma_{MB}(\mathbf{k},E)$ and $\Sigma^{\sigma<}_{MB}(\mathbf{k},E)$ to Eqs. (\ref{retardedGF}) and (\ref{lesserGF}) (see, for example, Ref. \onlinecite{bookStefanucci}).
%, which describe the feedback via interactions of the particle with surrounding electrons. 
Thus the Green's functions of the CR with interacting electrons become\cite{fe.ca.05,fe.ca.05_2,th.ru.08}
\begin{equation}\begin{split}
 G^\sigma(\mathbf{k}, E)=& [ES(\mathbf{k})-H^\sigma(\mathbf{k}) -\Sigma^\sigma_L(\mathbf{k},E)-\Sigma^\sigma_R(\mathbf{k},E)\\&-\Sigma^\sigma_{MB}(\mathbf{k},E)]^{-1},
  \end{split}\label{retardedGF_MB}
\end{equation}
and
 \begin{equation}\begin{split}
 G^{\sigma <}(\mathbf{k}, E)=&G^\sigma(\mathbf{k}, E)[\Sigma^{\sigma<}_L(\mathbf{k},E)+\Sigma^{\sigma<}_R(\mathbf{k},E)\\
 &+\Sigma^{\sigma<}_{MB}(\mathbf{k},E)]G^\sigma(\mathbf{k}, E)^\dagger.
 \end{split}
 \end{equation}
 Eq. (\ref{retardedGF_MB}) can be re-expressed as a Dyson equation
 \begin{equation}
    G^\sigma(\mathbf{k}, E)=g^\sigma(\mathbf{k}, E)+ g^\sigma(\mathbf{k}, E)\Sigma^\sigma_{MB}(\mathbf{k},E)G^\sigma(\mathbf{k}, E),\label{eq.Dyson}
\end{equation}
which allows to obtain the retarded many-body Green's function, also called dressed Green's function, from the non-interacting, or bare, Green's function $g^\sigma(\mathbf{k}, E)$ of Eq. (\ref{retardedGF}). \\
 The formal introduction of the many-body self-energies shows that the electron-electron interaction effectively acts on the system as an additional electrode.
 We can then define the effective ``coupling'' matrix\cite{dr.ru.17}
\begin{equation}
\Gamma_{MB}^\sigma(\mathbf{k},E)=i[\Sigma^\sigma_{MB}(\mathbf{k},E)-\Sigma^\sigma_{MB}(\mathbf{k},E)^\dagger],
\end{equation}
and express $\Sigma^<_{MB}(\mathbf{k},E)$ as\cite{dr.ru.17}
 \begin{equation}
  \Sigma^{\sigma<}_{MB}(\mathbf{k},E)=iF^\sigma_{MB}(\mathbf{k},E)\Gamma^\sigma_{MB}(\mathbf{k},E).
 \end{equation}
This equation has the same structure as Eq. (\ref{eq.Sigma_less}), but $F^\sigma_{MB}(\mathbf{k},E)$ is a matrix, which describes the out-of-equilibrium distribution of the interacting electrons in the CR, and it is not the Fermi function.
The resulting current was first computed by Meir and Wingreen in a seminal work\cite{me.wi.92}, and can be written as\cite{dr.ru.17} 
\begin{equation}
I= I_c+I_{nc}.\label{eq.current_MB}
\end{equation}
$I_c$ is the coherent contribution expressed as in Eq. (\ref{I_coherent}), 
but with the transmission coefficient $T^\sigma(E)$ evaluated with the retarded dressed Green's function, i.e., replacing $T^\sigma(\mathbf{k},E)$ with
\begin{equation}
  T^\sigma_{MB}(\mathbf{k},E)= \mathrm{Tr}\Big[\Gamma^\sigma_{L}(\mathbf{k},E)  G^\sigma(\mathbf{k}, E)^\dagger  \Gamma^\sigma_{R}(\mathbf{k},E) G^\sigma(\mathbf{k}, E)\Big].\label{eq.T_MB}
\end{equation}
$I_{nc}$ is the non-coherent contribution, and reads\cite{ne.da.10,dr.ru.17} 
\begin{equation}\begin{split}
  I_{nc}=
 &\sum_\sigma\mathrm{Tr}\Big\{[F^\sigma_{MB}(\mathbf{k},E)-f_R(E)] \times\\
 &\Gamma^\sigma_{MB}(\mathbf{k},E)  G^\sigma(\mathbf{k}, E)^\dagger  \Gamma^\sigma_{MB}(\mathbf{k},E) G^\sigma(\mathbf{k},E) \Big\}.\label{eq.I_nc}
 \end{split}
 \end{equation}
It accounts for an effective ``interaction electrode''. Electrons can be seen as entering the interaction electrode, where they undergo some scattering processes losing coherence, before being re-injected into the system\cite{fo.he.07}. The mathematical form of $I_{nc}$ resembles that of $I_c$. However, $F^\sigma_{MB}(\mathbf{k},E)$ cannot be brought outside the trace, and $I_{nc}$ cannot be associated to a transmission coefficient for the flow of electrons from the interaction electrode \cite{ne.da.10}. \\
Given Eq. (\ref{eq.current_MB}), the conductance $\mathfrak{G}$ can similarly be separated into a coherent and non-coherent contribution.
In the linear-response limit relevant for metallic heterostructures, the calculation of such non-coherent contribution is however an outstanding problem. To date, it has only been solved assuming either a specific shape for the matrix $F^\sigma_{MB}(\mathbf{k},E)$ (Refs. \onlinecite{fe.ca.05,ng.96}), or that the CR consists of a single orbital, so that $\Gamma^\sigma_{L/R}$ are numbers instead of matrices\cite{me.wi.92}. Alternatively, in the Kubo formalism, non-coherent contributions would be captured by vertex corrections\cite{og.01}, but we are not aware of any study, where these have been derived from first-principles calculations. 
Since the goal of our paper is not to provide a solution for this problem, but to present our implementation of LSDA+DMFT, we focus here on analysis of the transmission coefficient in Eq. (\ref{eq.T_MB}) rather than on the conductance. Importantly, $T^\sigma_{MB}(\mathbf{k},E_F)$ can be directly measured in experiment through the injection of hot electrons or holes in metallic heterostructures\cite{al.rz.17,ba.ha.05,ka.ro.08}.  Therefore, our calculations can provide physically relevant and verifiable predictions on correlation effects in quantum transport. This will be further discussed in Sec. \ref{Sec.spin_valve}.\\ 
In the zero-bias limit implied for metallic systems, $T^\sigma_{MB}(\mathbf{k},E_F)$ is calculated with the retarded dressed Green's function evaluated in thermodynamics equilibrium. 
%we skip over this issues and we focus on the linear-response limit. 
%Since our devices are composed of transition metals, which are described by Fermi liquid theory, we have $\Gamma^\sigma_{MB}(\mathbf{k},E_F)=0$. 
%Non-coherent contributions to the zero-bias conductance therefore vanish as inferred from Eq. (\ref{eq.I_nc})
%and the linear response conductance is expressed by Eq. (\ref{conductance}), but with $T^\sigma_{MB}(\mathbf{k},E_F)$ replacing $T^\sigma(\mathbf{k},E_F)$.
The fluctuation-dissipation theorem holds, and gives\cite{bookStefanucci}
\begin{equation}
G^{\sigma <}(\mathbf{k}, E)= i f(E)D^{\sigma }(\mathbf{k}, E),
 \end{equation}
where $f(E)=f_L(E)=f_R(E)$, and
\begin{equation} 
 D^{\sigma }(\mathbf{k}, E)=i[G^{\sigma}(\mathbf{k}, E)- G^{\sigma}(\mathbf{k}, E)^\dagger] \label{eq.D_MB}
  \end{equation}
is the spectral function. 
Thus, only $G^{\sigma}(\mathbf{k}, E)$ is required to fully describe the system. In summary, solving the interacting problem and calculating the transport properties within the mentioned approximations reduce to the evaluation of the retarded self-energy $\Sigma^\sigma_{MB}(\mathbf{k},E)$ and of the Dyson equation, Eq. (\ref{eq.Dyson}). 

 \subsection{Projection to the correlated subspace}\label{Sec.Projection1}
The discussion in the previous section provided formal equations to study transport in correlated nano-devices.
However, calculations including all the orbitals of the CR represent a great challenge in practical calculations.
To simplify the problem, we take advantage of the fact that in the case of TM-based heterostructures there are $4s$ and $3d$ valence states. 
$4s$ states are delocalized forming energy bands with a large dispersion, and electronic correlations are well described at the effective single-particle Kohn-Sham level. 
In contrast, open $3d$ shells are more tightly bound to the ionic cores, and, as such, they are moderately correlated. 
We then define the ``correlated subspace'' ($\mathcal{C}$) as the subspace of the CR that includes all $3d$ orbitals. 
Assuming that there are $N_{\mathrm{TM}}$ TM atoms inside the CR, the correlated subspace $\mathcal{C}$ has dimension $2(5\times N_{\mathrm{TM}})$ 
(the factor $2$ accounts for the spin).  
The correlated subspace can be projected out from the rest of the system, which we refer to as the “bath” (B), and which includes the orthogonal subspace to $\mathcal{C}$ within both the CR and electrodes.
To carry out such projection, we use the scheme presented in Ref.~\onlinecite{dr.ru.17}, and we perform the basis change, which transforms the CR overlap and Hamiltonian matrix as
\begin{equation}\begin{split}
\bar{S}(\mathbf{k})=\left( \begin{array}{cc}
 1 &   0          \\
 0 &   \bar{S}_\mathrm{B} (\mathbf{k})          \\
\end{array} \right)=\\
=W(\mathbf{k})^\dagger S(\mathbf{k}) W(\mathbf{k}), \end{split}
\label{eq:Saibath}
\end{equation}
\begin{equation}\begin{split}
\bar{H}^\sigma(\mathbf{k})=\left( \begin{array}{cc}
 \bar{H}^\sigma_\mathcal{C}(\mathbf{k}) &   \bar{H}^\sigma_{\mathcal{C},\mathrm{B}} (\mathbf{k})          \\
 \bar{H}^\sigma_{\mathrm{B},\mathcal{C}}(\mathbf{k}) &   \bar{H}^\sigma_\mathrm{B} (\mathbf{k})          \\
\end{array} \right)=\\
=W(\mathbf{k})^\dagger H^\sigma (\mathbf{k}) W(\mathbf{k}). \end{split}
\label{eq:Haibath}
\end{equation}
In the transformed matrices $\bar{S}(\mathbf{k})$ and $\bar{H}^\sigma(\mathbf{k})$
the top left block describes the correlated subspace $\mathcal{C}$, the bottom right
block describes the part of the bath included in the CR, and the off-diagonal blocks describe the connection terms.
The matrices $W(\mathbf{k})$ are defined in Eq. (10) of  Ref.~\onlinecite{dr.ru.17} .
The transformation is designed in such a way that the orbitals of $\mathcal{C}$ become orthogonal, and that they have zero overlap with the bath orbitals [see Eq. (\ref{eq:Saibath})].\\
$\bar{H}^\sigma_\mathcal{C}(\mathbf{k})$ in Eq. (\ref{eq:Haibath}) is
the non-interacting Hamiltonian matrix of $\mathcal{C}$ of dimension $5N_\mathrm{TM}\times5N_\mathrm{TM}$.
Using the second quantization formalism, we can introduce the Hamiltonian operator of $\mathcal{C}$
\begin{equation}
 \hat{\bar{H}}^\sigma_\mathcal{C}(\mathbf{k})=\sum_{i,j,\lambda_1, \lambda_2, \sigma}[\bar{ H}^\sigma_{\mathcal{C}}(\mathbf{k})]_{i \lambda_1, j \lambda_2} \hat d_{i \lambda_1 \sigma}^{\dag} \hat d_{j \lambda_2 \sigma},
\end{equation}
where $\hat d_{i \lambda \sigma}^{\dag}$ and
$\hat d_{i \lambda \sigma}$ are the electron creation and annihilation operators at orbital $\lambda$ within the atom $i$ and spin $\sigma$ ($i=1,...,N_\mathrm{TM}$ and $\lambda=1,...,5$, $\sigma=\uparrow, \downarrow$).
$[\bar{ H}^\sigma_{\mathcal{}}(\mathbf{k})]_{i \lambda_1, j \lambda_2}$ is the Hamiltonian matrix element
between the $d$ orbital $\lambda_1$ of the atom $i$ and the $d$ orbital $\lambda_2$ of the atom $j$.
Further, to describe the electron-electron interaction for the electrons in $\mathcal{C}$, we add an explicit Coulomb term as follows
\begin{equation}\begin{split}
& \hat{\bar{H}}^\sigma(\mathbf{k})_{\mathcal{C},U}= \hat{\bar{H}}^\sigma_\mathcal{C}(\mathbf{k})-\hat{H}^\sigma_{\mathcal{C},dc}+ \\ &+\frac{1}{2}\sum_{\substack{i,\lambda_1,\lambda_2,\lambda_3,\\ \lambda_4,\sigma_1,\sigma_2}}
U_{\lambda_1,\lambda_2,\lambda_3,\lambda_4} d_{i \lambda_1 \sigma_1}^{\dag} d_{i \lambda_2 \sigma_2}^{\dag} d_{i \lambda_4 \sigma_2} d_{i \lambda_3 \sigma_1}, \end{split}\label{Hint}
\end{equation}
where $U_{\lambda_1,\lambda_2,\lambda_3,\lambda_4}$ are the four-index Hubbard-$U$ matrix elements, which account for the screened Coulomb interaction between all $3d$ orbitals located on the same atom.
$U_{\lambda_1,\lambda_2,\lambda_3,\lambda_4}$ are parameterized in terms of the average effective Coulomb interaction $U$ and exchange $J$ (Ref. \onlinecite{pavarini})
\begin{eqnarray}
&U=\frac{1}{(2l+1)^2
}\sum_{\lambda_1,\lambda_2}U_{\lambda_1,\lambda_2,\lambda_1,\lambda_2}\label{U}\\
&J=\frac{1}{2l(2l+1) }\sum_{\lambda_1\neq
\lambda_2,\lambda_2}U_{\lambda_1,\lambda_2,\lambda_2,\lambda_1}.\label{J}
\end{eqnarray}
The reason for using the multi-orbital Hubbard-like form is
the local nature of the screened Coulomb interaction, which
allows us to ignore the Coulomb integrals involving correlated orbitals of different atoms.
$\hat{H}^\sigma_{\mathcal{C},dc}$ is the double-counting correction, which is needed to cancel the Coulomb interactions
already taken into account in the LSDA exchange-correlation potential.
The exact form of the double-counting correction is not known, but several approximations have been proposed and are used in practice \cite{li.ka.01,ko.sa.06,ka.ul.10,ha.ye.10}. 
We will describe our practical treatment of the problem in Secs. \ref{Sec.impurity} and \ref{sec.onsite}.\\ 
%From a pragmatic point of view we consider to include the double counting correction into the on-site energy shift as described in Sec. \ref{sec.onsite}.
We note that taking the static mean-field approximation for the Hubbard-like interaction in the Hamiltonian of Eq. (\ref{Hint}) leads to the so-called LSDA+$U$ method\cite{an.za.91,li.an.95,du.bo.98,co.gi.05}. This is one of the simplest corrective approaches that were formulated to improve the accuracy of LSDA functionals for correlated materials. We will further discuss it in Sec. \ref{Sec.impurity} and present some calculations in Appendix \ref{app.LDAU}. LSDA+$U$ has found widespread use for the computational design of functional materials. However, it was already shown in early works\cite{co.gi.05} that it can give worse results than LSDA for the electronic spectra and, therefore, the transport properties of the ferromagnetic metallic systems of our interest. To obtained improved results, the Hubbard-like interaction in $\hat{\bar{H}}^\sigma(\mathbf{k})_{\mathcal{C},U}$ needs to be treated beyond the static mean-field approximation introducing a many-body energy-dependent self-energy as we will describe in the rest of the paper.\\
 
 \subsection{Green's function and many-body self-energy of the correlated subspace}\label{Sec.Projection2}
 The bare and dressed Green's functions, $g^\sigma (\mathbf{k},E)$ and $G^\sigma (\mathbf{k},E)$, are expressed in the transformed basis as\cite{dr.ru.17}
 \begin{equation}\begin{split}
\bar{g}^\sigma(\mathbf{k},E)=\left( \begin{array}{cc}
 \bar{g}^\sigma_\mathcal{C}(\mathbf{k},E) &   \bar{g}^\sigma_{\mathrm{\mathcal{C},B}} (\mathbf{k},E)          \\
 \bar{g}^\sigma_\mathrm{B,\mathcal{C}}(\mathbf{k},E) &   \bar{g}^\sigma_\mathrm{B} (\mathbf{k},E)          \\
\end{array} \right)=\\
= W(\mathbf{k})^{-1 }g^\sigma (\mathbf{k},E) W(\mathbf{k})^{-1 \dagger}\end{split}
\label{eq:aibath}
\end{equation}
and
\begin{equation}\begin{split}
\bar{G}^\sigma(\mathbf{k},E)=\left( \begin{array}{cc}
 \bar{G}^\sigma_\mathcal{C}(\mathbf{k},E) &   \bar{G}^\sigma_{\mathrm{\mathcal{C},B}} (\mathbf{k},E) \\
 \bar{G}^\sigma_\mathrm{B,\mathcal{C}}(\mathbf{k},E) &   \bar{G}^\sigma_\mathrm{B} (\mathbf{k},E)          \\
\end{array} \right)=\\
= W(\mathbf{k})^{-1 }G^\sigma (\mathbf{k},E) W(\mathbf{k})^{-1 \dagger},\end{split}
\label{eq:dressedGaibath}
\end{equation}
 where $W(\mathbf{k})^{-1 }$ is the inverse of the transformation matrix used in Eqs. (\ref{eq:Saibath}) and (\ref{eq:Haibath}). 
 $\bar{g}^\sigma(\mathbf{k},E)$ and $\bar{G}^\sigma(\mathbf{k},E)$ have the same block structure as the Hamiltonian matrix in Eq. (\ref{eq:Haibath}). The blocks $\bar{g}^\sigma_\mathcal{C}(\mathbf{k},E)$ and $\bar{G}^\sigma_\mathcal{C}(\mathbf{k},E)$ are the bare and dressed Green's function matrices of the correlated subspace. They satisfy the Dyson equation for the correlated subspace 
\begin{equation}
\bar{G}^\sigma_{\mathcal{C}}(\mathbf{k}, E)=[\bar{g}^\sigma_\mathcal{C}(\mathbf{k}, E)^{-1}-\bar{\Sigma}^\sigma_{\mathcal{C}}(\mathbf{k},E)]^{-1}, \label{dressed_G_bar}
\end{equation}
where $\bar{\Sigma}^\sigma_{\mathcal{C}}(\mathbf{k},E)$ is the many-body self-energy of $\mathcal{C}$ and includes also the double counting correction. This self-energy formally specifies the electron correlations inside $\mathcal{C}$ due to the interaction in Eq. (\ref{Hint}). In the following sections we will see how $\bar{\Sigma}^\sigma_{\mathcal{C}}(\mathbf{k},E)$ is computed in practice.\\
From $\bar{\Sigma}^\sigma_{\mathcal{C}}(\mathbf{k},E)$ we can easily obtain the many-body self-energy of whole CR.  
 In the transformed basis, it reads
\begin{equation}
\bar{\Sigma}^\sigma_{MB}(\mathbf{k},E)=\left( \begin{array}{cc}
 \bar{\Sigma}^\sigma_\mathcal{C}(\mathbf{k},E) &   0    \\
 0 & 0         \\
\end{array} \right),    
\end{equation}
since the bath is non-interacting by construction. In the original basis the CR many-body self-energy is calculated as
 \begin{equation}
\Sigma^\sigma_{MB}(\mathbf{k},E)=W(\mathbf{k})^{-1^\dagger}
\bar{\Sigma}^\sigma_{MB}(\mathbf{k},E)
W(\mathbf{k})^{-1 }.\label{SigmaTrans}
\end{equation}
The many-body self-energy is therefore ``propagated'' to the bath.

\subsection{DMFT}\label{Sec.DMFT}
\begin{figure*}[ht]
\includegraphics[width=0.9\linewidth]{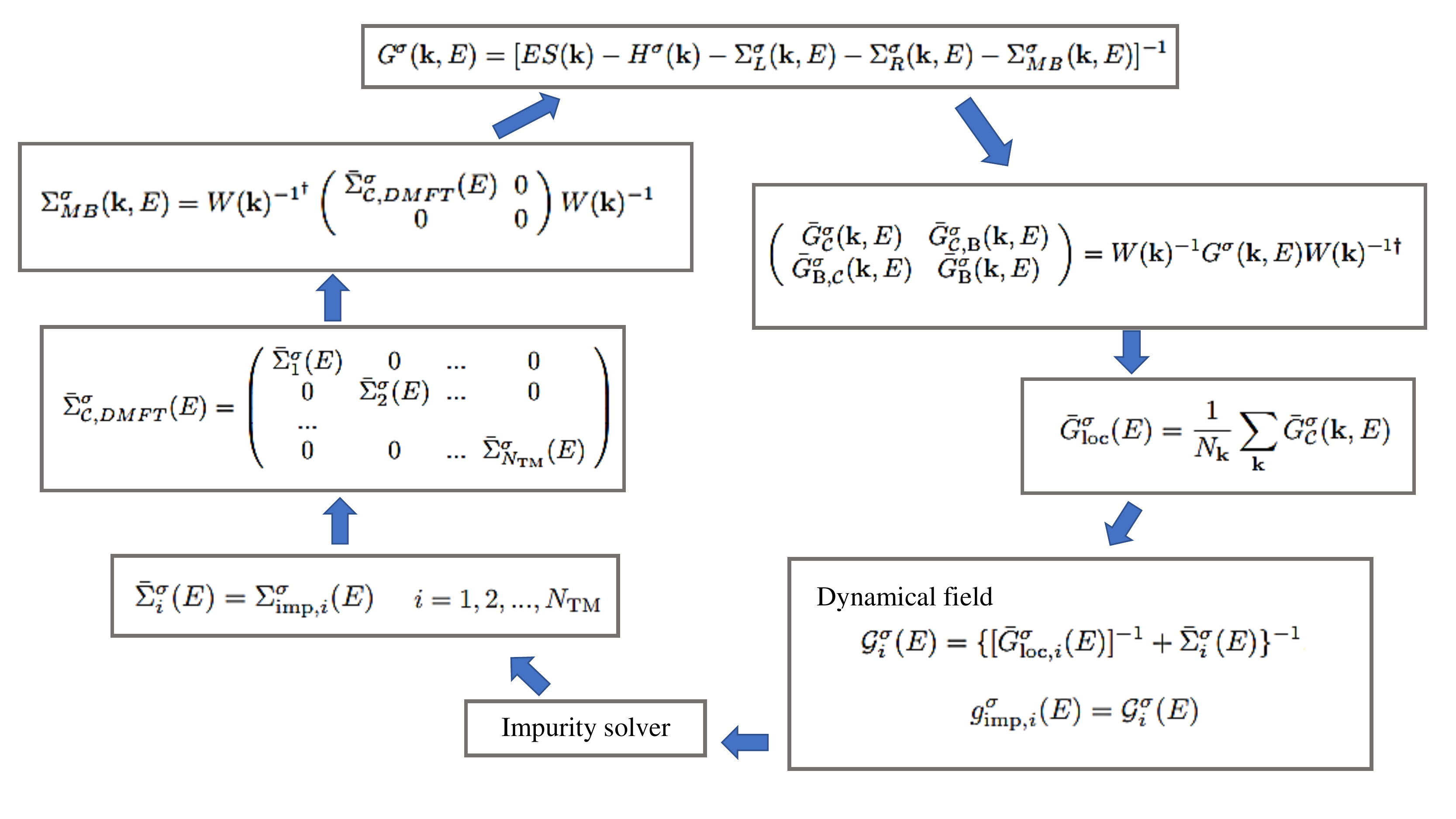}
\caption{Schematic representation of the DMFT self-consistent loop.}\label{Fig.DMFTloop}
\end{figure*}

The many-body problem within the correlated subspace $\mathcal{C}$ is solved via DMFT, which means that we only consider electron correlations local in space. The self-energy $\bar{\Sigma}^\sigma_\mathcal{C}(\mathbf{k},E)$ of $\mathcal{C}$ is therefore approximated by the DMFT self-energy
\begin{equation}
 \bar{\Sigma}^\sigma_{\mathcal{C},DMFT}(E)= 
 \left( \begin{array}{cccc}
 \bar{\Sigma}^\sigma_1(E) &   0 & ... & 0          \\
 0 & \bar{\Sigma}^\sigma_2(E) & ... & 0           \\
 0 & 0 & ... & \bar{\Sigma}^\sigma_{N_\mathrm{TM}}(E)            \\
\end{array} \right),
\label{Sigma_local}
\end{equation}
where $\bar{\Sigma}^\sigma_i(E)$ is the $5 \times 5$ block for the $3d$ orbitals of the atom $i$. We note that, in general, each block $\bar{\Sigma}^\sigma_i(E)$ can be non-diagonal. Eq. (\ref{Sigma_local}) generalizes to multi-orbital systems the DMFT self-energy used in layer-DMFT for tight-binding models \cite{ok.mi.04,po.no.99,va.sa.12}.  
Practically $\bar{\Sigma}^\sigma_{\mathcal{C},DMFT}(E)$ is computed by mapping the correlated subspace into a set of auxiliary impurity problems, one per atom. Each impurity problem is numerically solved obtaining the correspondent (local) many-body self-energies $\bar{\Sigma}^\sigma_i(E)$. The procedure is embedded into a self-consistency loop presented in Fig. \ref{Fig.DMFTloop}. %In Ref. \cite{dr.ra.arxiv}, we presented an implementation of the DMFT loop similar to that suggested by Valli {\it et al.} for model systems~\cite{va.sa.12} or Jacob {\it et al.} for nano-contacts~\cite{ja.ha.10}. The device was mapped into the correlated subspace, and the the DMFT self-consistent equations were converged within the CS. Here we present a different implementation.
Within our implementation, the CR is projected onto the correlated subspace at each DMFT iteration and, after solving the impurity problem, the self-energy is transformed back to the original space. By doing so, one retains more information about the states outside the correlated subspace. This is an approach that follows from previous implementations of LSDA+DMFT for periodic systems \cite{an.ko.05,le.ge.06,am.le.08,ai.po.09,pl.ha.18} and is here generalized to device setups. The self-consistent DMFT loop is summarized as follows:
\begin{itemize}
\item[i)] We compute the dressed CR Green's function $G^\sigma(\mathbf{k}, E)$ in Eq. (\ref{retardedGF_MB}). In the first iteration, we use $\Sigma^\sigma_{MB}(\mathbf{k},E)=0$ as a guess for the many-body self-energy.
\item[ii)] We carry out the transformation in Eq. (\ref{eq:dressedGaibath}), and we separate the Green's function of the correlated subspace $\bar{G}^\sigma_\mathcal{C}(\mathbf{k},E)$ from the rest of the system.   
\item[iii)]We define the so-called local Green's function 
\begin{equation}
  \bar{G}^\sigma_{\mathrm{loc}}(E)= \frac{1}{N_\mathbf{k}}\sum_\mathbf{k} \bar{G}^\sigma_{\mathcal{C}}(\mathbf{k}, E).\label{LocalGreen}
\end{equation}
\item[iv)] We build the $5\times5$ dynamical field matrix $\mathcal{G}^\sigma_{i}(E)$ for each atom $i$ in the correlated subspace:
\begin{equation}
\mathcal{G}^\sigma_{i}( E)=\{[\bar{G}^\sigma_{\mathrm{loc},i}( E)]^{-1}+ \bar{\Sigma}^\sigma_i(E)\}^{-1},\label{WeissField}
\end{equation}
where $\bar{G}^\sigma_{\mathrm{loc},i}( E)$ is the $5\times 5$ block of the local Green's function matrix relative to the atom $i$. 
\item[v)] We map each of the $N_\mathrm{TM}$ atom inside $\mathcal{C}$ into an Anderson impurity model. This is done by defining the bare impurity Green's function of atom $i$ as $g^\sigma_{\mathrm{imp},i}(E)=\mathcal{G}^\sigma_{i}( E)$.
\item[vi)] We solve the impurity problems as described in Sec.~\ref{Sec.impurity} thus obtaining the impurity many-body self-energies $\Sigma^\sigma_{\mathrm{imp},i}(E)$. 
\item[vii)] We set $\bar{\Sigma}^\sigma_i(E)=\Sigma^\sigma_{\mathrm{imp},i}(E)$ for each atom $i$, and we then compute the DMFT self-energy in Eq. (\ref{Sigma_local}).
\item[viii)] We transform back the self-energy to the original basis using Eq. (\ref{SigmaTrans}). This gives the updated CR many-body self-energy $\Sigma_{MB}^\sigma(\mathbf{k},E)$. We then go back to step i) to start the next iteration.  
\end{itemize}
After converging the self-consistent DMFT equations, we compute the density of states (DOS)
\begin{equation}
\mathrm{DOS}^\sigma(E)=\frac{1}{N_\mathbf{k}}\sum_\mathbf{k}D^\sigma(\mathbf{k},E),
\end{equation}
where $D^\sigma(\mathbf{k},E)$ is the spectral function defined in Eq. (\ref{eq.D_MB}), and the transmission coefficient $T^\sigma(E)=\frac{1}{N_{\mathbf{k}}} \sum_\mathbf{k} T_{MB}^\sigma(\mathbf{k},E)$ with $T_{MB}^\sigma(\mathbf{k},E)$ defined in Eq.(\ref{eq.T_MB}).
%The advantage of this implementation with respect to our other implementations of LSDA+DMFT proposed for transport \cite{ja.ha.10, dr.ra.arxiv} 
The transmission coefficient is, in principle, independent on the basis and can equivalently be computed in the original or in the transformed basis. However, it is more convenient to use the original basis thus employing the same module already implemented for DFT+NEGF in the Smeagol code.\\
We note that, in spite of the local approximation of DMFT, the self-energy in the original basis, $\Sigma_{MB}^\sigma(\mathbf{k},E)$, acquires a $\mathbf{k}$-dependence because of the matrices $W(\mathbf{k})^{-1}$ in Eq. (\ref{SigmaTrans}). The matrices $W(\mathbf{k})$ and their inverse are computed once at the beginning of the DMFT cycle and then stored. They are a basis transformation and basis orbitals do not change within the DMFT loop.
We would need to update them if we performed full charge self-consistent calculations including self-energy effects. Our implementation can in principle deal with these calculations, but in practice they remain too computationally demanding for real device setups. Moreover, we expect that charge self-consistency will have a minor impact for the metallic systems studied in this work as the main effect of correlation is to reduce the DOS spin-splitting, without altering the chemical bond and the charge distribution around the atoms.\\
Although in this paper we consider only moderately correlated systems, and the impurity solver is designed for them (see the next section), it is important to remark that our DMFT algorithm is general and also applicable to study nano-junctions comprising strongly correlated materials. In these cases, one would only need to opt for a different impurity solver, for example, employing the non-crossing~\cite{gr.ke.81,co.84} or the one-crossing approximations~\cite{ha.ki.01,pr.gr.89}. Any impurity solver can in principle be interfaced with our code in a straightforward way.\\

\subsection{Solution of the impurity problem}\label{Sec.impurity}
 As outlined in the previous section, DMFT requires the solution of auxiliary impurity problems to determine the self-energies $\bar{\Sigma}^\sigma_{i}(E)=\Sigma^\sigma_{\mathrm{imp},i}(E)$. The impurity solvers generally used for ferromagnetic metals, such as continuous time quantum Monte Carlo\cite{gu.mi.11} or the spin-polarized $T$-matrix fluctuating exchange approximation\cite{li.ka.98,ka.li.99,ka.li.02,po.ka.06}, are formulated on the imaginary frequency axis. Spectral functions are obtained indirectly via the numerical analytical continuation to the real energy axis\cite{ja.gu.96,sa.98,mi.pr.20,fu.pr.10}.
Unfortunately, this often leads to numerical difficulties, since the analytical continuation of discrete numerical data is not unambiguous. Moreover, it necessitates the appropriate treatment of the high-frequency ``tails''~\cite{se.he.91}. These issues become even more pressing in case of transport calculations, since the transmission coefficient in Eq. (\ref{eq.T_MB}) is computed from the retarded Green's functions, which may present many specific energy-dependent features. For this reason, here we consider the second order perturbative treatment proposed in Refs. \onlinecite{dr.ja.99,dr.ra.22}, and which is implemented to provide the self-energy directly on the real energy axis, while retaining the multi-orbital nature of the many-body problem. In spite of its simplicity, we showed in Ref. \onlinecite{dr.ra.22} that the second order self-energy already accounts for all characteristic spectroscopic features caused by electron correlation in ferromagnetic TMs.\\
The systems considered in this paper have diagonal dynamical field matrices $\mathcal{G}^\sigma_{i}( E)$ in Eq. (\ref{WeissField}) because of symmetry. This greatly reduces the computational effort for solving the impurity problem. The diagonal elements of the impurity Green's function and self-energy for an orbital $\lambda$ of an atom $i$ are denoted as $g^{\sigma}_{\mathrm{imp},i \lambda }(E)$ and $\Sigma^\sigma_{\mathrm{imp},i\lambda}(E)$. They are related through the impurity Dyson equation 
\begin{equation}
G^\sigma_{\mathrm{imp},i\lambda}(E)^{-1}=g^\sigma_{\mathrm{imp},i\lambda}(E)^{-1}-\Sigma^\sigma_{\mathrm{imp},i\lambda}(E),\label{eq.Dyson_impurity}
\end{equation}
with $G^\sigma_{\mathrm{imp},i\lambda}(E)$ the impurity dressed Green's function.\\
The self-energy up to the second order in diagrammatic perturbation theory in $U$ over the band width is written as
\begin{equation}\label{eq:sigma_1+2}
 \Sigma^\sigma_{\mathrm{imp},i \lambda }(E)\approx \Sigma^{\sigma (1)}_{\mathrm{imp},i \lambda }+\Sigma^{\sigma (2)}_{\mathrm{imp},i \lambda }(E)-\Sigma^{\sigma}_{dc,i \lambda }.
\end{equation}
$\Sigma^{\sigma}_{dc,i \lambda }$ represents the double counting correction, which we will discuss at the end of this section.
The first-order term is
\begin{equation}
\Sigma^{\sigma (1) }_{\mathrm{imp},i \lambda } = 
\sum_{\lambda_1 \sigma_1} U_{\lambda\lambda_1\lambda\lambda_1} n_{i \lambda_1}^{\sigma_1} - \sum_{\lambda_1} U_{\lambda\lambda_1\lambda_1\lambda} n_{i \lambda_1}^{\sigma},\label{eq.first}
\end{equation}
and is the well-known Hartree-Fock approximation, where $n_{i \lambda}^{\sigma}=\int_{-\infty}^\infty dE f(E) \textrm{Im}g^{\sigma}_{\mathrm{imp},i \lambda }(E)$
is the occupation of the orbital $\lambda$ of spin $\sigma$ for the impurity $i$; $f(E)$ is the Fermi function.
$\Sigma^{\sigma (1)}_{\mathrm{imp},i \lambda }$ is local in time, i.e. frequency independent. It therefore represents a one-electron potential producing only a shift of the non-interacting energy levels. In practice, if only $\Sigma^{\sigma (1)}_{\mathrm{imp},i \lambda }$ and the double counting correction were included in the calculations, this would correspond to using LSDA+$U$ instead of DMFT, as also mentioned at the end of Sec. \ref{Sec.Projection2}.\\
The second order self-energy $\Sigma^{\sigma (2)}_{\mathrm{imp},i \lambda }(E)$, which includes the dynamical correlations, can be split into its real and imaginary parts.
The imaginary part is given by~\cite{alig.06}
\begin{widetext}
\begin{eqnarray}
\label{eq:sig2r}
\mathrm{Im}\left[\Sigma^{\sigma (2)}_{\mathrm{imp},i \lambda}(E)\right] &= 
-\pi \sum\limits_{\lambda_1\lambda_2\lambda_3\sigma_1}& U_{\lambda\lambda_1\lambda_2\lambda_3}U_{\lambda_3\lambda_2\lambda_1\lambda} 
\int_{-\infty}^\infty \mathrm{d}\epsilon_1 \int_{-\infty}^\infty\mathrm{d} \epsilon_2 
D_{i\lambda_1}^{\sigma_1}(\epsilon_1) 
D_{i\lambda_2}^{\sigma}(\epsilon_2) 
D_{i\lambda_3}^{\sigma_1}(\epsilon_1+\epsilon_2-E) \times \nonumber \\
&&\{ f(\epsilon_1) f(\epsilon_2) + 
\left[ 1-f(\epsilon_1) - f(\epsilon_2) \right] f(\epsilon_1+\epsilon_2-E) \} \nonumber \\
&+\pi \sum\limits_{\lambda_1\lambda_2\lambda_3}& U_{\lambda\lambda_1\lambda_2\lambda_3}U_{\lambda_2\lambda_3\lambda_1\lambda}
\int_{-\infty}^\infty \mathrm{d}\epsilon_1 \int_{-\infty}^\infty\mathrm{d} \epsilon_2 
D_{i\lambda_1}^{\sigma}(\epsilon_1+\epsilon_2-E) 
D_{i\lambda_2}^{\sigma}(\epsilon_2) 
D_{i\lambda_3}^{\sigma}(\epsilon_1) \times \nonumber  \\
&&\{ f(\epsilon_2) f(\epsilon_1) + 
\left[ 1-f(\epsilon_2) - f(\epsilon_1) \right] f(\epsilon_1+\epsilon_2-E) \},\label{eq.second}
\end{eqnarray}
\end{widetext}
 where
 \begin{equation}
D_{i \lambda}^\sigma(E) = -\frac{1}{\pi} \mathrm{Im}[g_{\mathrm{imp},i \lambda}^{\sigma }(E) ] \label{eq.D_i}
\end{equation}
is the spectral function of $g_{\mathrm{imp},i \lambda}(E)$.
The real part is computed by the Kramers-Kronig relations as
\begin{equation}
    \mathrm{Re} \left[ \Sigma^{\sigma (2)}_{\mathrm{imp},i \lambda}(E) \right] = -\frac{1}{\pi} \int_{-\infty}^{\infty} d\epsilon \frac{\mathrm{Im}\left[\Sigma^{\sigma (2)}_{\mathrm{imp},i \lambda}(\epsilon)\right]}{E-\epsilon}.\label{eq.KK}
\end{equation}
 We note that the dressed rather than the bare impurity Green's function could be used in the evaluation of self-energy contributions\cite{bookStefanucci}. This would require a self-consistent solution of Eqs. (\ref{eq.Dyson_impurity}), (\ref{eq.first}), and (\ref{eq.second}). In this paper, we do not consider this approach to reduce the computational cost of the calculations. Effectively, we neglect some of the second order diagrams in the perturbative expansion\cite{bookStefanucci}, and, therefore, some multi-band screening effects. However, these effects are expected to be small\cite{dr.ra.22}, and hence not significant for the goals of this paper. \\
Our calculations in Sec. \ref{Sec.Results} are practically performed as follows. We approximate the first order and double counting contributions of the self-energy with the static potential of the LSDA+$U$ formulation by Dudarev {\it et al.}~\cite{du.bo.98}
\begin{equation}
\Sigma^{\sigma (1)}_{\mathrm{imp},i \lambda }-\Sigma^{\sigma}_{dc,i \lambda} \approx V_{U,i\lambda}^\sigma=(U-J)(\frac{1}{2}-n_{i\lambda}^\sigma).\label{eq.Dudarev}
\end{equation}
$V_{U,i\lambda}^\sigma$ is obtained through a charge self-consistent calculation. Then this static potential is included into the bare Green's function. This means that
$g^\sigma_{i \lambda }( E)$ is replaced by the LSDA+$U$ Green's function 
\begin{equation}
 g^\sigma_{\mathrm{LSDA}+U,i \lambda }( E)= [g^\sigma_{\mathrm{imp},i \lambda }( \omega)^{-1}-V_{U,i\lambda}^\sigma]^{-1}.\label{GF_LDAU}
\end{equation}
The Dyson equation, Eq. (\ref{eq.Dyson_impurity}), 
retains its structure, but the total self-energy $\Sigma^\sigma_{\mathrm{imp},i \lambda}(E)$ is substituted by the correlation self-energy $\Sigma^\sigma_{\mathrm{corr},i \lambda}(E)=\Sigma^\sigma_{\mathrm{imp},i \lambda }(E)- V_{U,i\lambda}^\sigma$, which is evaluated using $g^\sigma_{\mathrm{LSDA}+U,i \lambda }( E)$ in Eq. (\ref{eq.second}). Further details can be found in Ref. \onlinecite{dr.ra.22}, where the performances of the method for ferromagnetic TMs are also assessed against the results of photomoemission spectroscopy experiments.\\

\subsection{On-site energy shift}\label{sec.onsite}
DFT+NEGF calculations are performed in the gran-canonical ensemble. Here, rather than fixing the total number of electrons inside the CR, the Fermi energy is fixed by the chemical potential of the electrodes, $\mu_L=\mu_R=E_F$. The total number of electrons in the CR typically fluctuates during the charge self-consistent DFT cycle until it eventually converges to the nominal value given by the sum of the various CR atomic valence (+core) electrons for pseudopotentials-based (all-electron) implementations. A similar behaviour is expected also in charge self-consistent LSDA+DMFT two-terminal device calculations.
However, these charge self-consistent LSDA+DMFT calculations are computationally too demanding for realistic systems like those studied here. In line with typical calculations for periodic systems, we therefore perform self-energy self-consistent DMFT calculations, but we do not iterate the evaluation of the charge density. The total number of electrons of the CR is found to deviate slightly from the nominal value. To reimpose the correct electron counting for periodic systems, one usually adjusts the chemical potential of the impurity until the correct occupation is obtained. We adapt this process to the transport setup by adding an on-site atom-dependent potential $v_i$ to all correlated $3d$ orbitals. In other words, we readjust the real part of the many self-energy as $\mathrm{Re}\big[\Sigma^\sigma_{\mathrm{imp},i \lambda }(E)\big]\rightarrow \mathrm{Re}\big[\Sigma^\sigma_{\mathrm{imp},i \lambda }(E)]+v_i$. 
%The potential $v$ incorporates an effective double-counting correction for two-terminal device setups. 
We note that this is an ad-hoc adjustment based on %no other argument than
the electron counting. Yet, preliminary studies, albeit for different and simpler systems, seem to suggest that such adjustment reproduces quite well the results of fully charge self-consistent calculations\cite{dr.ru.unpublished}.
%To our best knowledge the formulation of a specific double-counting scheme for a two-terminal device has not yet been implemented and this does not constitute the goal of the present study. 
%This interesting problem, which, to our knowledge, has never been raised before, along with a choice for a double counting correction suitable for device simulations, goes beyond the goal of this paper and is left for future studies.   

\section{Computational details}\label{Sec.Details}
DFT calculations are performed treating core electrons with norm-conserving Troullier-Martin pseudopotentials\cite{tr.ma.91}. 
The valence states are expanded through a numerical atomic orbital basis set including multiple-$\zeta$ and polarized functions~\cite{so.ar.02}. 
The electronic temperature is $300$ K.
The real space mesh is set by an equivalent energy cutoff of $300$ Ry. A $\mathbf{k}$-point mesh equal to $k_x\times k_y=24\times24$ is used to compute the self-consistent charge density with DFT. 
This charge density is then used as input in a non-self-consistent DFT calculation with $80\times80$ $\mathbf{k}$-points to obtain the DOS. All energies are shifted in such a way to set the Fermi level at 0 eV. The Cu lattice constant is set to the experimental value, 3.615~\AA, and we do not optimize the structures.\\
DMFT calculations are performed using the bare Green's function  $g(\mathbf k,E)$ of Eq. (\ref{eq:aibath}) calculated for $32\times32$ $\mathbf{k}$-points. The temperature is $300$ K.
An energy grid comprising $3200$ points and extending from $-16$ to $6$ eV is employed to  calculate the second-order self-energy. The Coulomb parameters $U_{\lambda_1,\lambda_2,\lambda_3,\lambda_4}$ are expressed in terms of Slater integrals $F^0$, $F^2$ and $F^4$ (Ref.~\onlinecite{im.fu.98}). 
These are connected to the average effective Coulomb and exchange interactions of Eqs. (\ref{U}) and (\ref{J}) through the relations $U=F^0$ and 
$J=(F^2+F^4)/14$. The ratio $F^4/F^2$ is assumed to correspond to the atomic value $\approx 0.625$ (Ref.\onlinecite{an.gu.91}).

\begin{figure}[h]
\centering\includegraphics[width=0.55\textwidth,clip=true]{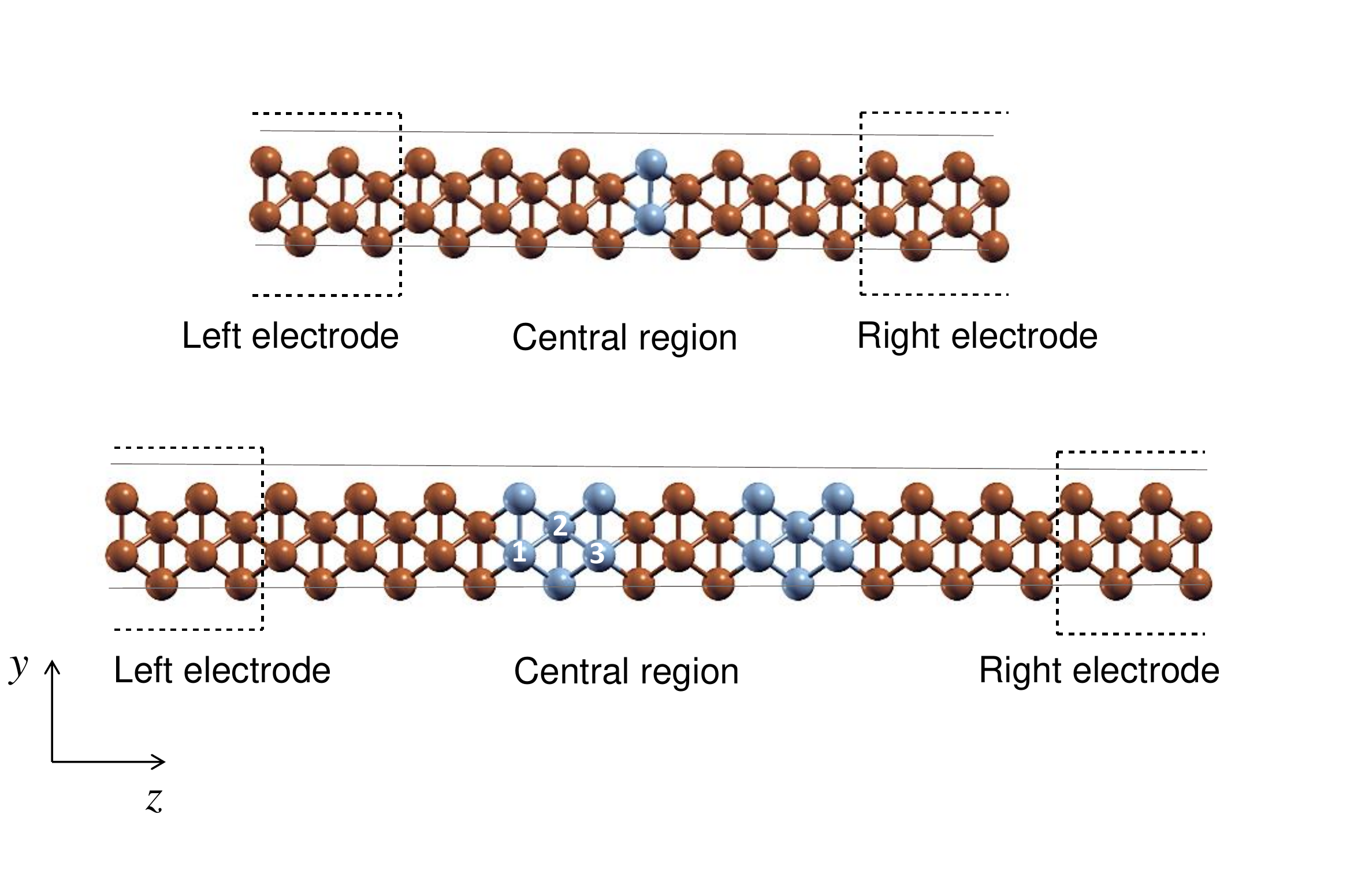}
\caption{Cu/Co/Cu (top panel) and Cu/Co$_3$/Cu$_3$/Co$_3$/Cu (bottom panel) two-terminal devices. The atoms Co1, Co2 and Co3 in  Cu/Co$_3$/Cu$_3$/Co$_3$/Cu are labelled in the figure.}
\label{pic.CoCuCo}
\end{figure}

\section{Results}\label{Sec.Results}
We now apply the method to heterostructures presenting alternating Cu and Co layers, sandwiched between semi-infinite Cu electrodes. The goal is to illustrate the capabilities of our implementation of LSDA+DMFT and, in doing so, to gain some general understanding about the impact of electron correlation effects on the electronic structure and transport properties.  
We first consider a single Co layer
and then a more complex heterostructure, whose central region comprises two Co trilayers separated by a Cu spacer, and which display GMR effect. \\
LSDA+DMFT is systematically compared to DFT within the LSDA, which is the standard theoretical approach considered in previous works about Co and Cu heterostructures (for example, see Refs. \onlinecite{sc.ke.95,sc.vh.97,bu.zh.95,xi.ke.01}). 
For completeness and to further compare static vs dynamic mean field effects, we also present some results of LSDA+$U$ calculations in appendix \ref{app.LDAU}. \\
The correlated subspace includes only the Co $3d$ orbitals, while the Cu $3d$ orbitals are considered uncorrelated, since they are fully filled and located in energy at about $2$ eV below the Fermi level. We use the four-index interaction term as shown in Eq. (\ref{Hint}) with the average Coulomb and exchange interactions set to $U= 3.0 $ eV and $J=0.9$ eV. These are standard values for Co, and we obtain a DOS similar to that obtained in  calculations \cite{ch.mo.15} based on the exact muffin-tin orbitals (EMTO)-DMFT method\cite{an.sa.00,vi.01,vi.sk.00,ch.vi.03}. In appendix \ref{app.DMFT_U_J} we present a systematic analysis of the dependence of the results on the interaction parameters.
Finally, the potential $v_i$ mentioned in Sec. \ref{sec.onsite}, and added to maintain the nominal charge of the central region, is set to be same for all Co atoms as they are found to be almost equivalent.

\begin{figure}[h]
\centering\includegraphics[width=0.48\textwidth,clip=true]{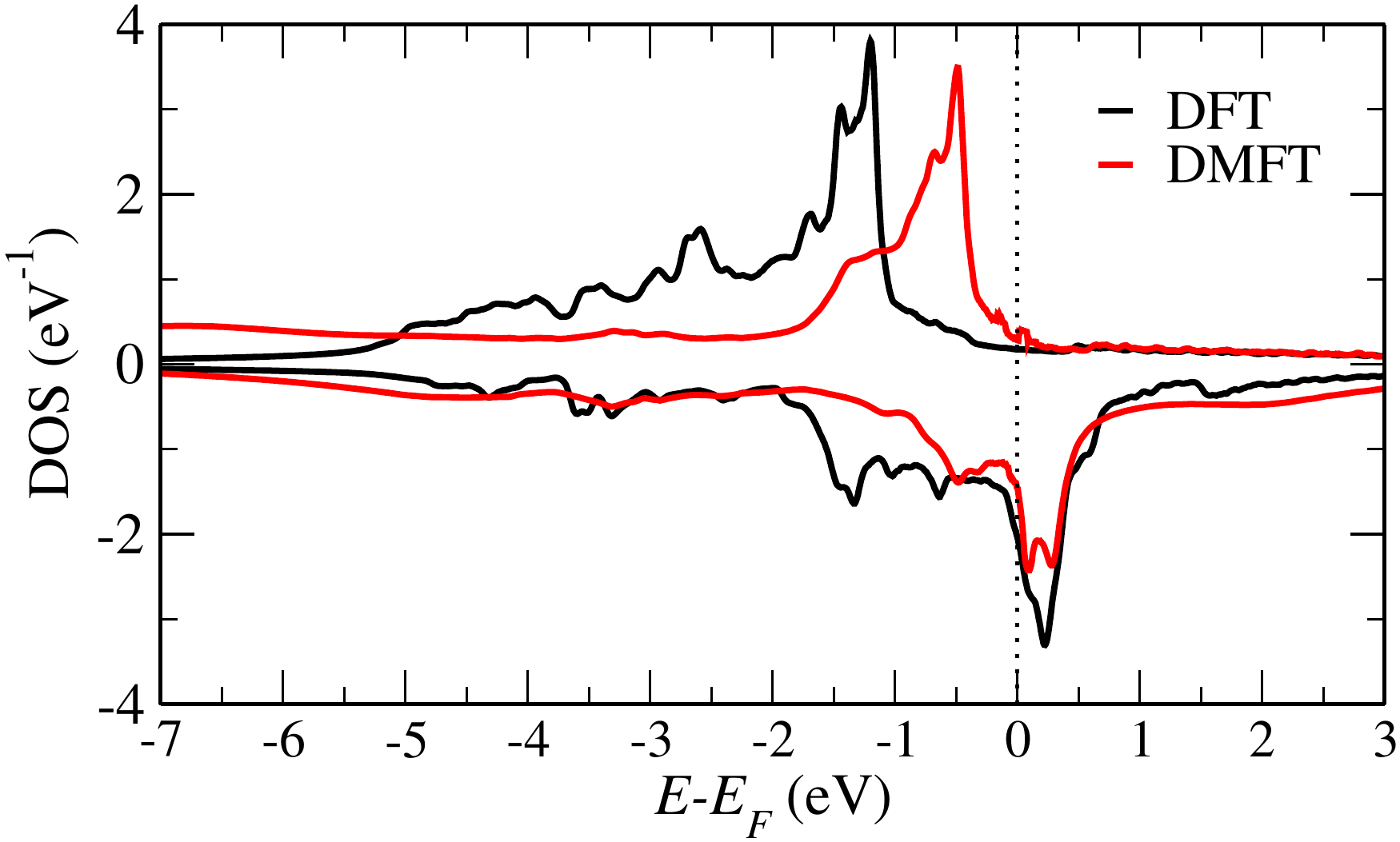}
\caption{DOS of the Co layer in the Cu/Co/Cu system, calculated by using DFT and DMFT.}
\label{DOS1Co}
\end{figure}

\subsection{Correlated Co monolayer}\label{Sec.monolayer}

\begin{figure}[h]
\centering\includegraphics[width=0.48\textwidth,clip=true]{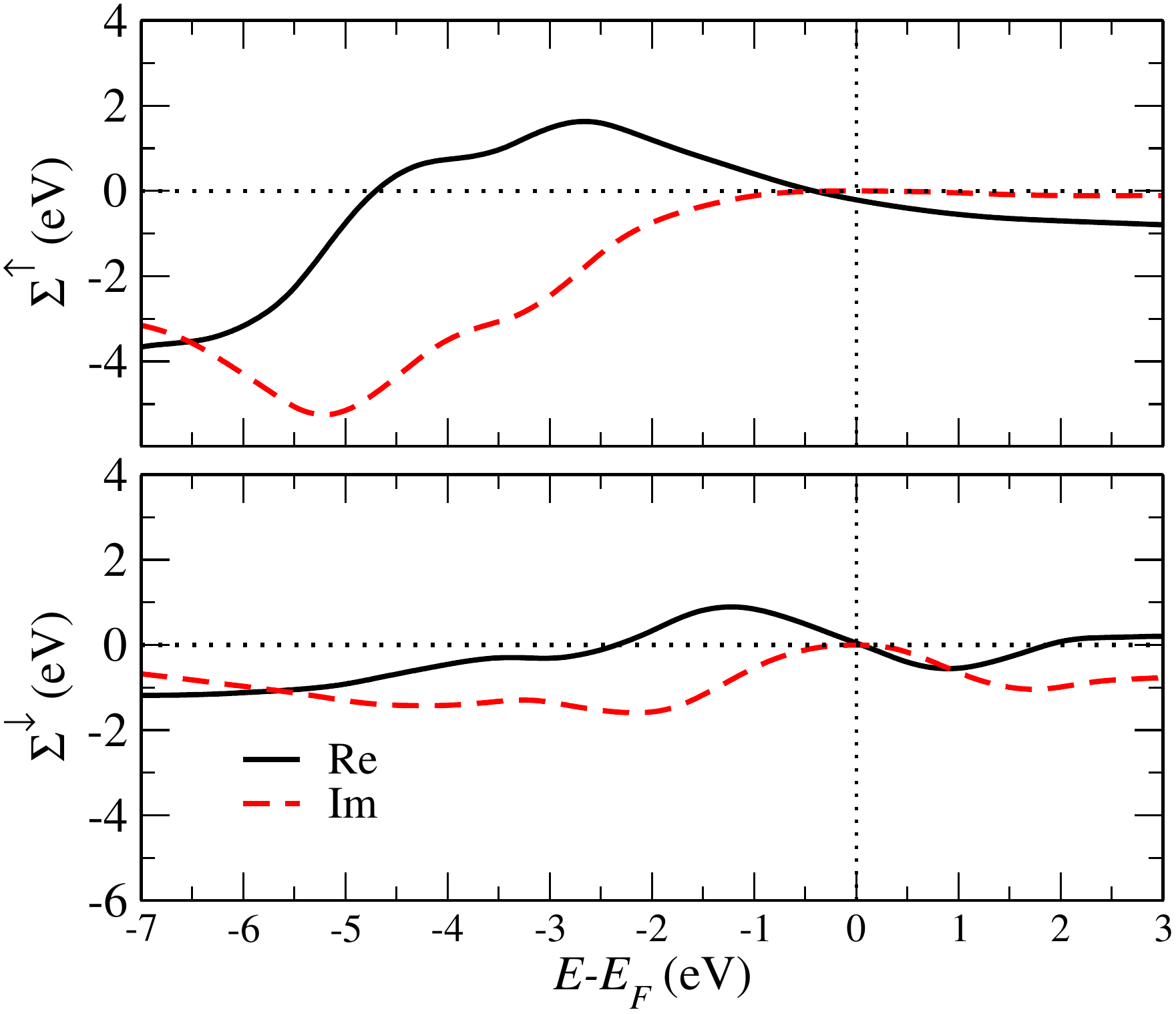}
\caption{Real part (solid black line) and imaginary part (dashed red line) of the many-body self-energy averaged over all Co $3d$ orbitals in the Cu/Co/Cu system. The upper panel is for majority (spin up), while the bottom panel is for minority (spin down) electrons.}
\label{Fig.Sigma}
\end{figure}

\begin{figure}[h]
\centering\includegraphics[width=0.47\textwidth,clip=true]{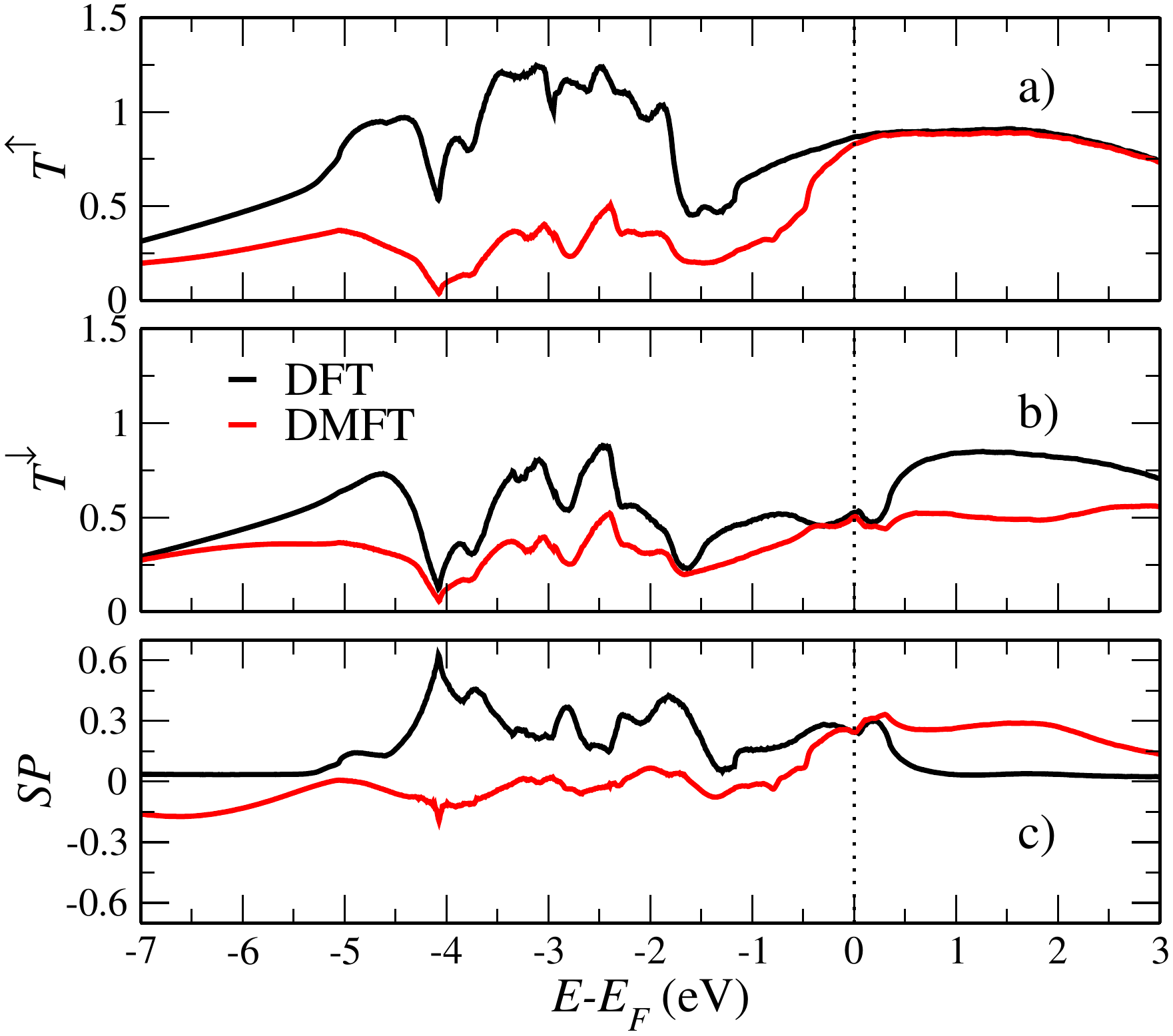}
\caption{(a) Transmission coefficient as a function of energy for the Cu/Co/Cu system, for majority (spin up) electrons. (b) Transmission coefficient as a function of energy for minority (spin down) electrons. (c) Spin-polarization as a function of energy. DFT (DMFT) results are in black (red).}
\label{fig.TRC1Co}
\end{figure}

We denote the system as Cu/Co/Cu. The simulation cell is shown in top panel of Fig. \ref{pic.CoCuCo}. The transport direction $z$ is oriented along the Cu(001) direction.
The DOS of the Co monolayer is shown in Fig. \ref{DOS1Co}. It is similar to that of bulk Co showing a strong ferromagnetic character \cite{Li.Pa17}. 
In the LSDA results,
the majority (spin up) $d$-states are almost fully occupied, and they are split by about $1.5$ eV from the minority (spin down) states, which cut through the Fermi level. A small spin-polarization is also induced via hybridization on the Cu layers in proximity to Co as discussed in Appendix \ref{app.DOS_Cu}. 
In the DMFT calculations, the dynamical self-energy induces a redistribution of the spectral weight. 
The changes in the DMFT DOS with respect to the DFT DOS are more pronounced for the majority than for the minority channel. The majority $3d$ states are shifted towards the Fermi level, while the position of the minority states is barely affected. As a result, the spin splitting is reduced by about $0.6$ eV compared to the DFT value, and it becomes equal to $0.9$ eV. Beside that, the DOS is considerably narrowed for energies close to the Fermi level, while it broadens below $E-E_F\approx -3$ eV. Overall, these changes are typical for correlation effects in TMs. As discussed in a number of works (for example Refs. \onlinecite{li.ka.01,br.mi.06,gr.ma.07,ch.mo.15}) DMFT accurately captures them. In particular, the good performance of our implementation with the perturbative solver is demonstrated in Ref. \onlinecite{dr.ra.22} through a comparison against photoemission spectroscopy experiments.\\
The changes in the DOS due to dynamical correlation effects are understood by inspecting the many-body self-energy. To simplify the analysis, we express it in the transformed basis, Eq. (\ref{Sigma_local}), noting that the main features in the Co DOS look the same in the original and transformed basis. We further take the average over the orbital indexes $\Sigma^\sigma(E)=\sum_\lambda\Sigma^\sigma_\lambda(E)/5$ to point those general features, which will be important later when analyzing the transmission coefficient (the overbar above the self-energy symbol used in Sec. \ref{Sec.DMFT} is neglected here to make the notation lighter).
$\Sigma^\sigma(E)$ is shown in Fig.~\ref{Fig.Sigma} 
It has Fermi-liquid character near the Fermi level: the imaginary part goes to zero as $-\mathrm{Im}\Sigma^\sigma(E)\propto (E-E_F)^2$. Away from the Fermi energy, it is much larger for the majority than for the minority channel, indicating that the
majority electrons are more correlated than the minority electrons.
The absolute magnitude of $\mathrm{Im}\Sigma^\uparrow(E)$ grows for $E-E_F \lesssim-2$~eV resulting in the substantial broadening of the DOS in that energy range as already seen in Fig. \ref{DOS1Co}.
$\mathrm{Re}\Sigma^\uparrow(E)$ shows a maximum at $E-E_F \approx-2.5$ eV. This causes the large shift in energy of the $3d$ states from the DFT position towards the Fermi level discussed above.
In the minority channel, $\mathrm{Re}\Sigma^\downarrow(E)$ has a peak below the Fermi energy, specifically at $E-E_F\approx-1.2$ eV. The $3d$ states, in particular $3d_{x^2-y^2}$, at that energy are therefore moved from their DFT position towards the Fermi level. For negative energies, the absolute value of $\mathrm{Im}\Sigma^\downarrow(E)$ is considerably smaller than that of $\mathrm{Im}\Sigma^\uparrow(E)$. Minority $3d$ states are indeed much less broadened than majority states in the energy region far below the Fermi level as distinctly seen in the DOS in Fig. \ref{DOS1Co}. In contrast, $\mathrm{Im}\Sigma^\downarrow(E)$ becomes quite large above the Fermi level, in particular for $E-E_F \gtrsim 1$~eV. This feature will have an impact on the transmission coefficient in that energy region, and it will also be important for the results of the next section.\\
%\begin{figure}[h]
%\centering\includegraphics[width=0.48\textwidth,clip=true]{Sigma_d.pdf}
%\caption{Real and imaginary part of the many-body self-energy for the Cu/Co/Cu system, for majority (spin up) and minority (spin down) electrons, averaged over the Co $d$ states.}
%\label{Fig.Sigma}
%\end{figure}
The spin-dependent DFT and DMFT transmission $T^\sigma(E)$, obtained respectively using Eqs. (\ref{eq.TRC}) and (\ref{eq.T_MB}), is depicted in Fig. \ref{fig.TRC1Co}. In the DFT calculations the transmission is quite large ($> 0.5$) over the whole displayed energy range, because the system is an all-metal heterostructure. %$T^\sigma (E)$ presents quite sharp peaks in the energy region below the Fermi level, where the Co $3d$ states are located, while it is quite smooth above the Fermi energy, where there are only $s$ states.
On the other hand, in the DMFT calculations, the transmission is drastically suppressed. This is because of two effects dominant at different energies. First, the Co $3d$ states acquire a finite relaxation time $\tau$, which is related to the imaginary part of the many-body self-energy, $\tau^{-1} \propto \mathrm{Im}\Sigma^\sigma$. Second, the occupied Co $3d$ states are dragged towards the Fermi level by the real part of the self-energy. As a result of that, the $s$ conduction electrons undergo a more pronounced elastic scattering at the Co layer in the DMFT picture than in the DFT picture.
Focusing in particular on the majority spin channel (Fig. \ref{fig.TRC1Co}-a), we note that $T^\uparrow (E)$ calculated with DFT presents quite sharp peaks in the energy region between $E-E_F\approx -5$ and $-1.3$ eV, where the $3d$ states are located. These peaks are suppressed in the DMFT transmission, mostly because of the finite relaxation time. In contrast, at energies from $E-E_F\approx -1.3$ eV to $1$ eV, where there are the $s$ states, and $\mathrm{Im}\Sigma^\uparrow(E)$ is very small, the DMFT majority transmission $T^\uparrow (E)$ is reduced compared to the DFT one because of the elastic scattering of the conduction $s$ electrons with the Co $3d$ orbitals. In the minority spin channel, this effect is less important as the energy position of the 
$3d$ states is not drastically modified by DMFT (see Fig. \ref{fig.TRC1Co}-b). Yet, we observe a suppression of the transmission $T^\downarrow (E)$ in the two energy regions $-1.8\lesssim E-E_F< 0$ eV and $0< E-E_F\lesssim 2$ eV due to the finite relaxation time. The transmission right at the Fermi energy remains however almost unaffected since $\mathrm{Im}\Sigma^\downarrow(E_F)= 0$ due to the Fermi liquid nature of the system and $\mathrm{Re}\Sigma^\downarrow(E_F)\approx 0$.\\

Using the transmission coefficients we can now compute the energy-dependent spin-polarization
\begin{equation}
SP(E)=\frac{ T^{\uparrow}(E)-T^{\downarrow}(E)}{ T^{\uparrow}(E)+T^{\downarrow}(E)}.\label{eq.SP}
\end{equation}
The results are plotted in Fig. \ref{fig.TRC1Co}-c. DFT gives a positive spin-polarization at all energies below the Fermi level, and the largest values are found in the range from $E-E_F \approx -4$ to $-2$ eV, where the majority Co $3d$ states are located. In contrast, DMFT predicts that the spin-polarization in that energy region becomes very small and negative as a consequence of the suppression of the transmission in the majority channel. Near the Fermi energy, the DFT and DMFT spin-polarization are almost identical and equal to $0.24$ as both $T^{\uparrow}(E)$ and $T^{\downarrow}(E)$ are hardly affected by dynamical correlations.
The most interesting energy region is for $E-E_F \gtrsim 0.5$ eV. Here the DFT spin-polarization vanishes, but the DMFT one is quite large and positive. This is because, as seen above, $T^{\uparrow}(E)$ is due only to uncorrelated $s$ electrons and remains large in both DFT and DMFT, whereas $T^{\downarrow}(E)$ has a contribution from the minority $d$ states, which acquire a finite relaxation time in DMFT owing the significant imaginary part of the self-energy, and therefore gets partly suppressed. This important feature was already anticipated above when discussing the shape of $\mathrm{Im}\Sigma^\downarrow(E)$, and it will have further interesting implications for the GMR effect studied in the next section.

\begin{figure}[h]
\centering\includegraphics[width=0.48\textwidth,clip=true]{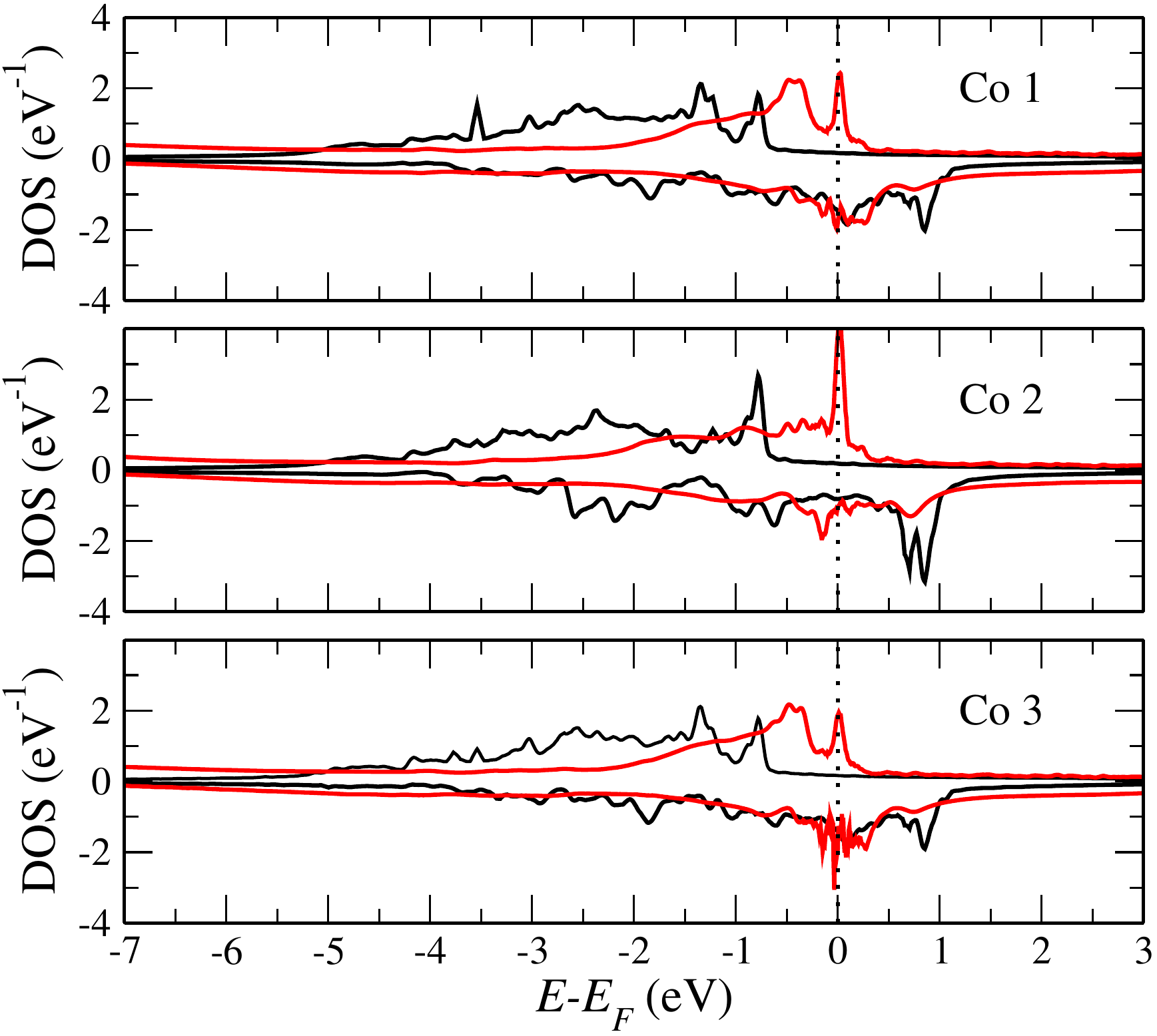}
\caption{DOS of the Co atoms in three different layers of the Cu/Co$_3$/Cu$_3$/Co$_3$/Cu system, calculated with DFT and DMFT. The atoms Co1, Co2 and Co3 are indicated in Fig. \ref{pic.CoCuCo}.}
\label{fig.DOS_3Co}
\end{figure}

%\begin{figure}[h]
%\centering\includegraphics[width=0.48\textwidth,clip=true]{orbital_DOS_central_Co.pdf}
%\caption{DOS projected over the $3d$ orbitals of the Co 2 atom in Cu/Co$_3$/Cu$_3$/Co$_3$/Cu. Top panel: DFT %calculation, Bottom panel: DMFT calculation.}
%\label{fig.DOS_central_Co}
%\end{figure}

\begin{figure*}[ht!]
\centering\includegraphics[width=0.8\textwidth,clip=true]{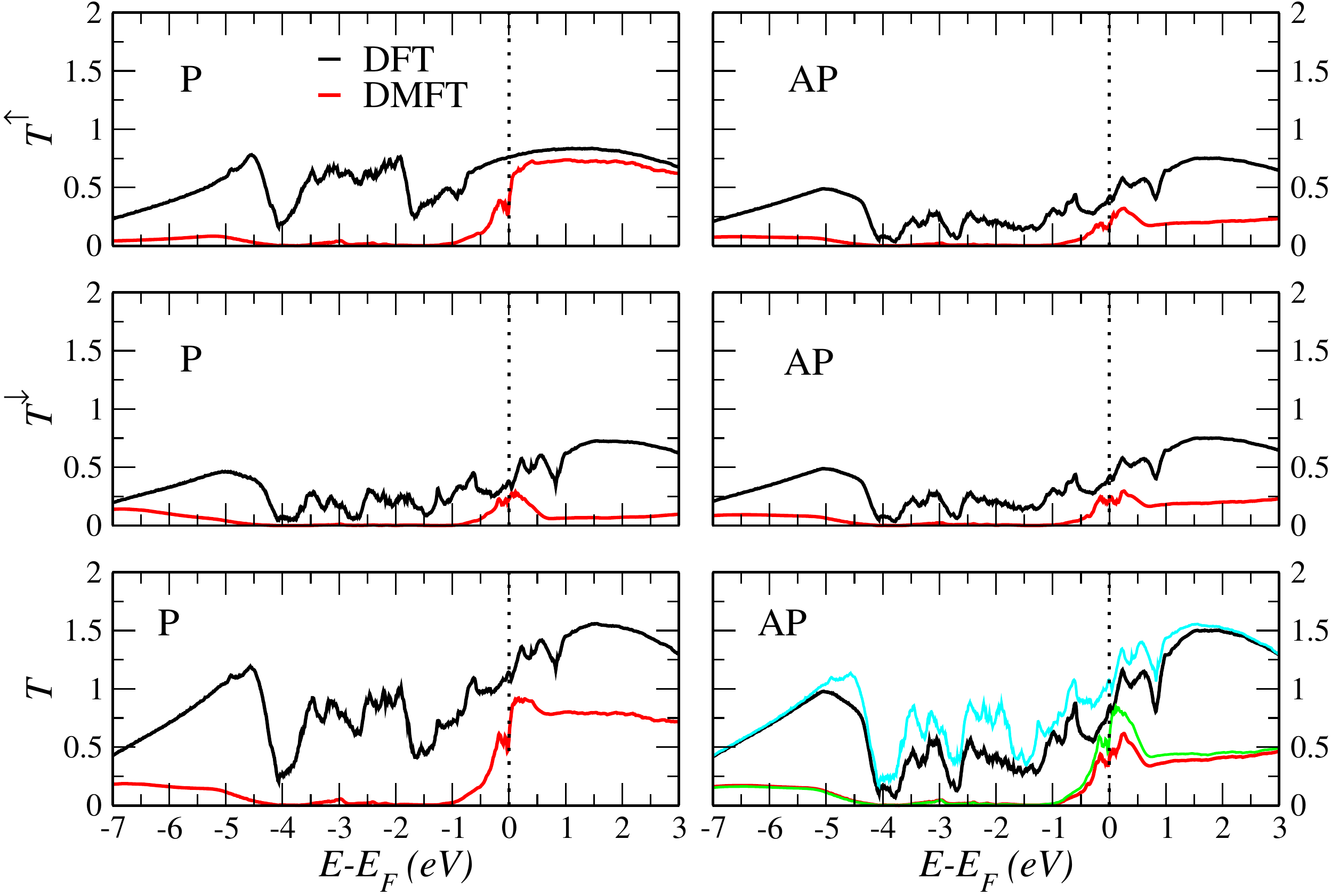}
\caption{Spin up, spin down and total transmission coefficient, $T^\uparrow(E)$, $T^\downarrow(E)$ and $T(E)=T^\uparrow(E)+T^\downarrow(E)$, for Cu/Co$_3$/Cu$_3$/Co$_3$/Cu (left panels: P configuration; right panels: AP configuration). In the bottom right panel, the cyan (green) curve represents the AP transmission coefficient calculated using the model approximation $T_{AP}(E) \sim 2\sqrt{T_P^\uparrow(E) T_P^\downarrow(E)}$ with the DFT (DMFT) $T_P^\uparrow(E)$ and $T_P^\downarrow(E)$.}
\label{TRC_GMR}
\end{figure*}

\subsection{Correlated Co multilayered device}\label{Sec.spin_valve}
\begin{figure}[h]
\centering\includegraphics[width=\columnwidth,clip=true]{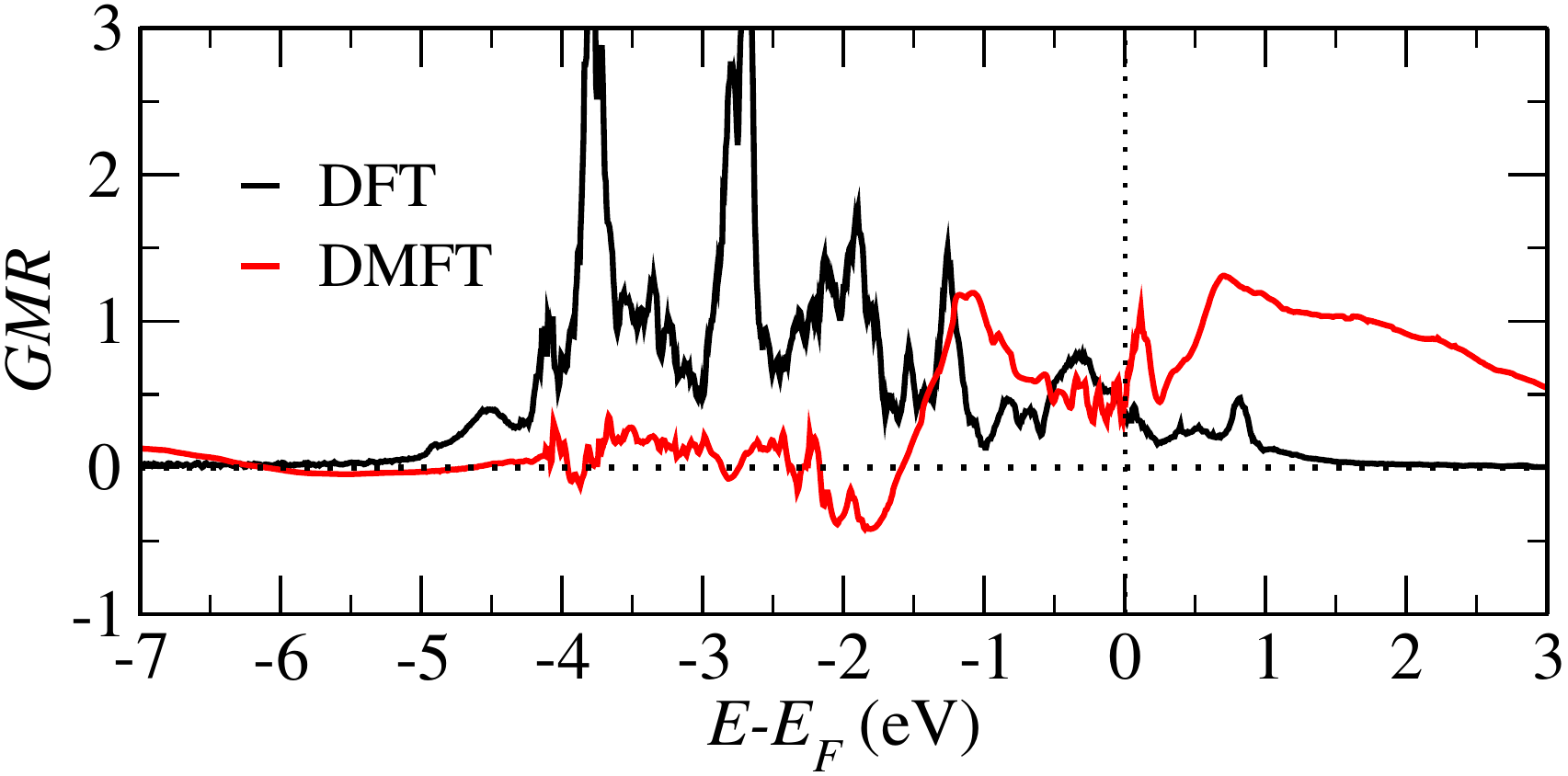}
\caption{Energy-dependent $GMR$ calculated by DFT (black line) and DMFT (red line).}
\label{fig.GMR}
\end{figure}

We consider here the heterostructure shown in the bottom panel of Fig. \ref{pic.CoCuCo} and named Cu/Co$_3$/Cu$_3$/Co$_3$/Cu, whose central region comprises two Co trilayers separated via a Cu spacer. It represents a spin-valve, where the magnetization of the first Co trilayer on the left side of the central region can be set parallel (P) or antiparallel (AP) to the magnetization of the second Co trilayer on the right side.\\
The DOS of one of trilayers is presented in Fig. \ref{fig.DOS_3Co}, where Co1, Co2 and Co3 label the non-equivalent Co atoms in the three different layers (see Fig. \ref{pic.CoCuCo}). % Co1 and Co3, which are at the interface with Cu, have almost identical electronic structure. Their DOS is very similar to that of the single Co layer analyzed in the previous section. In contrast, the DOS of Co2, which is located at the center of the trilayer, is significantly different. Co2 forms bonding and anti-bonding states with the surrounding Co atoms.  
%In the DFT calculations, the majority bonding and antibonding states extend over the energy regions $-5\lesssim E-E_F< -2.5$ eV and $-1.8\lesssim E-E_F< -0.5$ eV, respectively. The majority $d$ states are therefore almost fully occupied. In the minority channel, the bonding states are located in energy between $E-E_F \approx =-2.5$ and $-1$ eV, while the antibonding states cross the Fermi level. The crystal field splitting for Co2 is larger than for Co1 and Co3. Specifically, we  find a separation of about 0.5 eV between the triplet states, $d_{xy}$, $d_{xz}$ and $d_{yz}$, and the doublet state, $d_{x^2-y^2}$ and $d_{z^2}$. 
For all of them, DMFT induces a redistribution of the majority spectral weight compared to DFT, thus reducing the spin splitting of the $3d$ states. The effect of electron correlations is therefore the same as described in the previous section for the single Co layer, and the many-body self-energy (not shown) has a similar shape. \\
The spin-dependent transmission coefficients $T^\sigma_P(E)$ and $T^\sigma_{AP}(E)$ for the P and the AP magnetic configurations are shown in Fig. \ref{TRC_GMR}. 
The most striking feature is the suppression of the transmission through the Co $3d$ states predicted by DMFT regardless of the Co trilayers' magnetic alignment. The effect is more dramatic here than in the Co monolayer case. $T^\sigma_{P(AP)}(E)$ is of the order of 0.01 in the energy region between $ E-E_F \approx -4$ eV and $\approx-2$ eV in Cu/Co$_3$/Cu$_3$/Co$_3$/Cu, while it remained as large as about 0.25 in Cu/Co/Cu.
%We first analyze the P case. Both $T^\uparrow_P(E)$ and $T^\downarrow_P(E)$ are qualitatively very similar to the transmission obtained for Cu/Co/Cu. Nonetheless, we note that the suppression of the transmission through the Co $d$ states predicted by DMFT is more dramatic here than in the previous system. $T^\sigma_P(E)$ becomes of the order of 0.01 in the energy region between $ E-E_F \approx -4$ eV and $\approx-2$ eV, while it remained as large as about 0.25 in Cu/Co/Cu.
The cause is the presence of six Co layers instead of just one. Electrons acquire a finite relaxation time in each layer because of the imaginary part of the many-body self-energy. If there were more Co layers, the transmission would drop further. Similar results are found for both spin channels.
Overall, our analysis demonstrates that DMFT quantitatively captures the reduction of the coherent transmission due to electron relaxation. \\
%However, it is important to emphasize that incoherent contributions to the conductance are neglected in our calculations. 
%The incoherent contributions to the conductance are expected to become larger as the coherent transmission decreases. However, the inclusion of inelastic contributions to the non-equilibrium transport properties of complex systems unfortunately remains an open issue s outlined in Sec. \ref{Sec.Transport.Corr}.\\
We now analyze in more detail the P magnetic configuration. Both $T^\uparrow_P(E)$ and $T^\downarrow_P(E)$ are qualitatively very similar to the transmission coefficients previously obtained for Cu/Co/Cu.
In the majority channel, the conduction at energies $E-E_F\gtrsim -0.5$ is due to $s$ electrons. They are nearly free in DFT, whereas DMFT predicts a large scattering with the Co $3d$ states. These correlated states are placed close to the Fermi energy by the real part of the self-energy causing the reduction of the transmission. At the Fermi level, DMFT gives
$T^\uparrow_P(E_F)=0.33$, which is half the DFT result, $0.76$. 
In the minority channel, the reduction of the transmission coefficient at the Fermi level is slightly smaller. %Although the transport is mostly through $d$ states, correlation effects are less marked than in the majority channel. 
We obtain that $T^\downarrow_P(E_F)$ is equal to 0.22 and 0.35 in DFT and DMFT, respectively. 
%In the minority channel, the transport is through the $3d$ states. $T^\downarrow_P(E_F)$ is estimated equal to 0.22 and 0.35 in DFT and DMFT, respectively. 
On the other hand, above the Fermi energy, $T^\downarrow_P(E)$ is fully suppressed in DMFT. This is because the imaginary part of the spin down self-energy is quite large, as we highlighted in previous section. Thus we find that the total P transmission is $T_{P}(E)=T^\uparrow_{P}(E)+T^\downarrow_{P}(E)\approx  T^\uparrow_{P}(E)$. \\
In the AP configuration an electron belonging to the majority
band in the left Co trilayer will belong to the minority in the right trilayer, and vice-versa \cite{sanvito}. The transmission coefficient for spin up and spin down is identical,  $T^\downarrow_{AP}(E)=T^\uparrow_{AP}(E)$.
Spin up (down) electrons incoming from the left electrode with energies near $E_F$ go through the majority $s$ (minority $3d$) states of the left Co trilayer as well as the minority $3d$ (majority $s$) states of the right Co trilayer. Because of the mismatch between the $s$ and $3d$ states, electrons undergo a large elastic scattering in the central region, and the AP total transmission $T_{AP}(E)=T^\downarrow_{AP}(E)+T^\uparrow_{AP}(E)$ is significantly reduced compared to the P total transmission $T_{P}(E)$.
%In the energy region near and above the Fermi energy, majority $s$ electron incoming from the left trilayer will have to go through the minority $3d$ states of the right Co trilayer.%, thus undergoing a large scattering. As a result, $T^\uparrow_{AP}(E)$ is reduced compared to the P case.
%In the very same way incoming minority electrons will have to go through the minority $3d$ states of the left trilayer and the majority $s$ state of the right trilayer so that we find $T^\downarrow_{AP}(E)=T^\uparrow_{AP}(E)$. 
This physics is already captured at the qualitative level with DFT as shown in early works \cite{sc.ke.95}. However, our calculations indicate that there is a further drastic suppression of the transmission due to electron relaxation in the $3d$ states. This is evident when comparing the DFT and DMFT $T_{AP}(E)$ for $E-E_F>-0.5$ eV in Fig. \ref{TRC_GMR}. Overall, the dynamical self-energy contribution leads to a total P transmission $T_P(E)$, which is twice as large as the AP one $T_{AP}(E)$.\\ 
To better understand the impact of electron correlations on the transmission, we model the left Co/Cu and the right Cu/Co interfaces as two independent scatterers in series. We can then use the phenomenological expression for the AP transmission in terms of the spin up and spin down P  transmission, $T_{AP}(E) \sim 2\sqrt{T_P^\uparrow(E) T_P^\downarrow(E)}$ (Ref. \onlinecite{book1}). The results are represented in the bottom right panel of Fig. \ref{TRC_GMR} as the cyan and the green lines for DFT and DMFT, respectively. In the DMFT case, the model accurately describes $T_{AP}(E)$ for negative energies up to $E-E_F\approx$ -0.3 eV. This indicates that quantum interference effects in that energy region are largely suppressed by DMFT as a results of the imaginary part of the many-body self-energy.\\
Finally, we quantitatively characterize the spin transport properties of the system by computing the GMR ratio as a function of the energy  
\begin{equation}
 GMR(E)=\frac{ T_{P}(E)-T_{AP}(E)}{\mathrm{min}[T_{P}(E),T_{AP}(E)]},\label{eq.GMR}
\end{equation}
where $\mathrm{min}[T_{P}(E),T_{AP}(E)]$ indicates the smallest transmission coefficient between $T_P(E)$ and $T_{AP}(E)$ at the energy $E$. The results are shown in Fig. \ref{fig.GMR}. For negative energies, $GMR(E)$ is very large in DFT, whereas it is negligible in DMFT, since both $T_{P}(E)$ and $T_{AP}(E)$ are suppressed by electron relaxation. On the other hand, DMFT predicts a significant $GMR(E)$ for energies $E-E_F\gtrsim -0.5$ eV, where the P transmission is associated to spin up $s$ electrons, and, as such, is large and unaffected by correlation, while the AP transmission is reduced as both spin up and spin down electrons travel through the $3d$ states of either Co trilayers. This is a very interesting finding with important implications as discussed below.\\
The linear magnetoresistive response that is generally considered in the literature\cite{Sa.La.99,sc.ke.95} is obtained by evaluating Eq. (\ref{eq.GMR}) at the Fermi energy. DMFT predicts an increment of about $30\%$ with respect to DFT. Specifically, we get $GMR(E_F)=0.45$ in DMFT, and $GMR(E_F)=0.35$ in DFT. However, we must point out that this result is very sensitive to the interaction parameters $U$ and $J$ as we discuss in detail in Appendix \ref{app.DMFT_U_J}. Despite that, we observe that, for this specific system, DMFT always predicts an enhancement of $GMR(E_F)$ compared to the DFT. \\ 
A direct comparison of the linear response transport properties of metallic heterostructures with experiments has generally been  difficult\cite{xi.ke.01,sc.ke.97}. In most devices, the metallic layers are very thick ($\sim 100$ nm), and the physics is dominated by diffuse scattering and disorder\cite{ba.pr.99}, which cover any other phenomena, and which cannot be easily included in first-principles computational approaches. However, more recently, the direct measurement of transmission through materials has become possible in hot electron or hole transport experiments \cite{al.rz.17,ba.ha.05,ka.ro.08}. In particular, Kaidatzis {\it et al.} studied Co/Cu heterostructures using the ballistic electron emission microscope technique \cite{ka.ro.08}. By injecting electrons from a metallic tip into high energy Co/Cu conduction states, the authors could extract the spin-dependent transmission coefficients and, therefore, compute $GMR(E)$ as defined in Eq. (\ref{eq.GMR}) for a wide energy range above $E_F$. One of their most interesting results is that $GMR(E)$ is as large as about $2$ at $E-E_F\approx 1$ eV. It then decreases monotonically with energy, but nonetheless remains significant and equal to about 0.5 at $E-E_F\approx 2$ eV. This behaviour is clearly well captured, at least at the qualitative level, by the DMFT results of Fig. \ref{fig.GMR}. $GMR(E)$ presents a maximum at $E-E_F\approx 1$ eV (albeit equal to about 1.2 rather than 2) and falls off towards $0.5$ for larger energies in agreement with the experiments. On the hand, the effect is totally absent in the DFT results, where $GMR(E)$ is seen to rapidly drop just above $E_F$ thus becoming negligible at large energies. 
Based on these observations, we propose that the hot electron GMR is a clear manifestation of dynamical correlation effects. Our study demonstrates the potential of our implementation of DMFT for understanding new physics in metallic heterostructures.

\section{Conclusions}
We present a computational scheme, which uses LSDA+DMFT in combination with the NEGF approach to investigate spin-dependent electron transport through nanoscale two-terminal devices including electron correlation effects. We consider second-order perturbation theory for the impurity solver, which allows us to compute the many-body self-energy directly for real energies, thus avoiding numerical problems due to the analytic continuation. Perturbation theory is only appropriate for moderately correlated systems, such as $3d$ ferromagnetic TMs. However, our code can be easily interfaced with other impurity solvers to treat also strongly correlated materials. LSDA+DMFT calculations with our perturbative solver are only slightly more complicated and computational demanding than standard DFT+NEGF calculations, thereby making our method an ideal tool for the wider user community.\\
We apply our LSDA+DMFT method to heterostructures comprising alternating Co and Cu layers to obtain their zero-bias coherent transmission coefficient. We find that such transmission is suppressed by electron correlations at energies away from $E_F$. This is due to the finite imaginary part of the many-body self-energy, which corresponds to the inverse of an effective electron lifetime. In contrast, at $E_F$, the imaginary part of the self-energy vanishes due to the Fermi liquid behaviour, so that the changes in the transmission are entirely determined by the correlation-induced shift of the energy spectrum. In particular, the elastic scattering of uncorrelated majority $s$ electrons with the Co $3d$ states can be enhanced in DMFT compared to DFT. In some cases, this can greatly affect the linear-response spin-dependent coherent conductance. \\
The calculated transmission coefficient as a function of the energy can be used to interpret hot electron transport experiments. In particular, based on our LSDA+DMFT results, we suggest that the GMR measured in Cu/Co heterostructures for electrons with energies 1 eV-larger than $E_F$ is a peculiar manifestation of dynamical correlation effects. Encouraged by our study, we believe that LSDA+DMFT will soon help to find other many-electron features in quantum transport experiments.
%As a result, the GMR of spin-valve devices is reduced compared to DFT estimates. In the specific system studied here this reduction is of about 17\%. Since the magnitude of the change in GMR is dictated by the DMFT-induced shifts of the electronic structure, in general arbitrarily large changes in GMR will be possible if the shift of peaks induced by DMFT is large.
%Our results do not aim at explaining experiments, but to illustrate the performances of our implementation of DMFT. Nonetheless, the prediction that electron correlation reduces the GMR is important.The lower values of the GMR obtained in experiments compared to estimates based on DFT-NEGF are usually attributed to interface roughness, impurities etc. Although these are indeed very important and not negligible, here we argue that electron correlation may also give a contribute. We plan to investigate in details this problem in future studies.   
\section{Acknowledgements}
AD acknowledges funding by the Science Foundation Ireland (SFI) and the Royal Society through the University Research Fellowship URF-R1-191769. 
MMR acknowledges funding provided by the Institute of Physics
Belgrade, through the grant by the Ministry of Education,
Science, and Technological Development of the Republic of
Serbia. 
IR acknowledges the support of the UK government department for Business, Energy and Industrial Strategy through the UK national quantum technologies
programme.
LC acknowledges the financial support by the Deutsche Forschungsgemeinschaft through TRR80 (project F6) Project number 107745057. 
Computational resources were provided by Trinity College Dublin Research IT. 

\appendix
\section{LSDA+$U$ calculations}\label{app.LDAU}

\begin{figure}[h]
\centering\includegraphics[width=\columnwidth,clip=true]{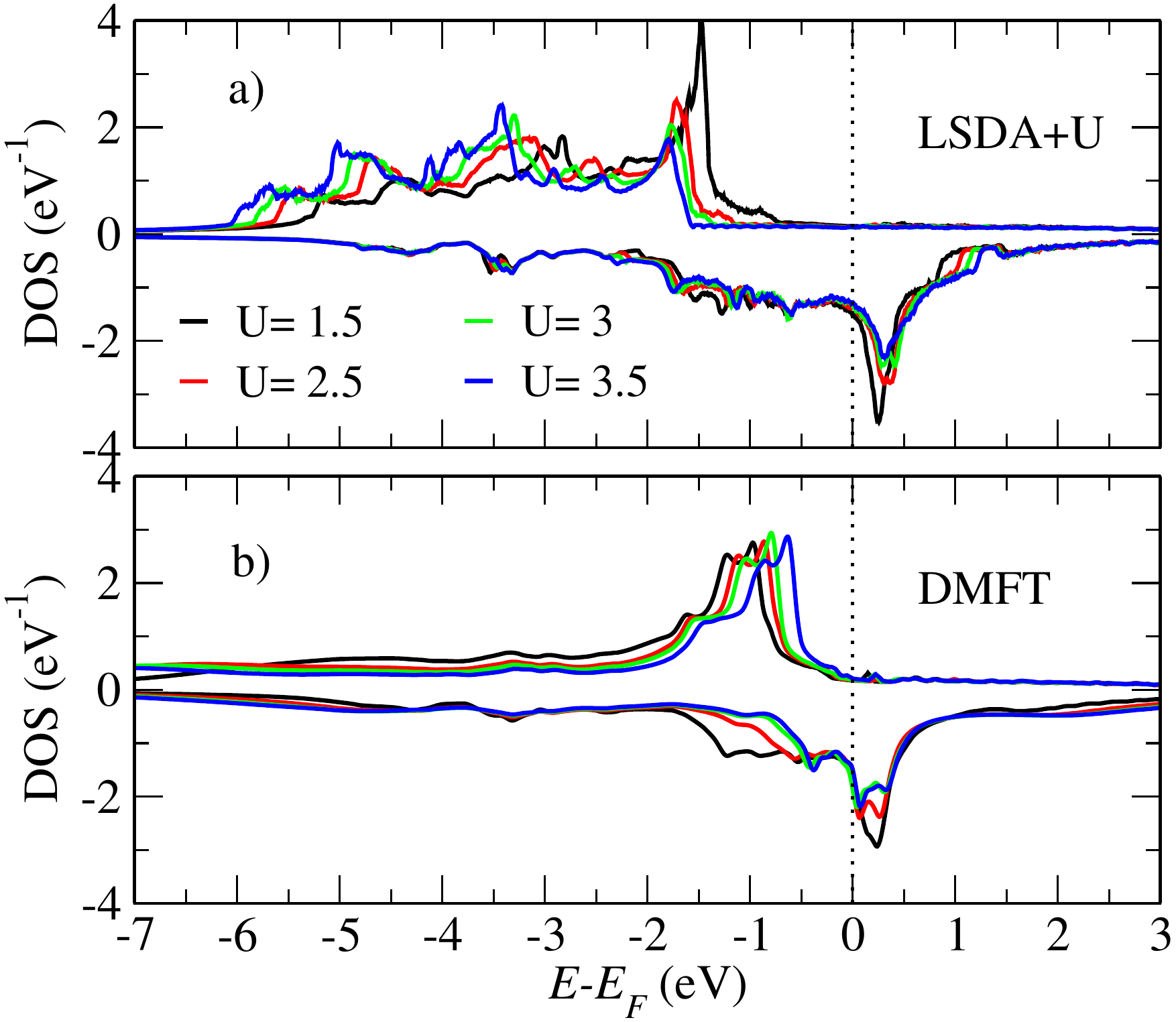}
\caption{DOS of the Co layer in the Cu/Co/Cu system, calculated by using LSDA+$U$ and DMFT for $1.5$ eV (black line), $2.5$ eV (red line), $3$ eV (green line) and $3.5$ eV (blue line). $J$ is constant and equal to $0.5$ eV.}
\label{fig.DOS_U_J0.5}
\end{figure}
We present here the electronic and transport properties of Cu/Co/Cu calculated by using the LSDA+$U$ approach. We consider the formulation by Dudarev {\it et al.}~\cite{du.bo.98} introduced in Sec. \ref{Sec.impurity}. 
The inspection of results help to grasp the impact of dynamical over static mean-field approximations.
The DOS of the Co atom is shown in Fig. \ref{fig.DOS_U_J0.5}-a for $U$ equal to $1.5$ eV, $2.5$ eV, $3$ eV and $3.5$ eV, while $J$ is fixed at 0.5 eV.
Since the majority states are almost fully filled, the Hubbard-like mean-field corrective potential, which is defined in Eq.~(\ref{eq.Dudarev}), drags the spin up LSDA DOS towards lower energies by about $-0.5(U-J)$.
In contrast, in the minority channel, the effect of the potential is less marked. The main peak in the DOS, which is due to the $d_{xy}$ orbital above the Fermi level, is moved by about $0.1(U-J)$ towards high energies. Overall, the DOS spin splitting becomes larger when increasing $U$.  
This general behaviour was already found in early DFT+$U$ calculations for ferromagnetic TMs~\cite{co.gi.05}.
%As a result, the spin-splitting of the bands is even more overestimated than in LSDA calculations.
%and it is due to the treatment of the Hubbard interaction within the static mean-field approximation. 
As already discussed in our previous work\cite{dr.ra.22}, the $U$ static potential actually worsen, instead of improving, the capability of DFT to reproduce the electronic spectra of $3d$ TMs leading to a drastic overestimation of the DOS spin-splitting. Dynamical correlations are needed to correct that.\\ %The correlation captured by the dynamic self-energy plays a crucial role in counterbalancing the static mean-field approximation of LSDA+$U$. \\

\begin{figure}[h]
\centering\includegraphics[width=\columnwidth,clip=true]{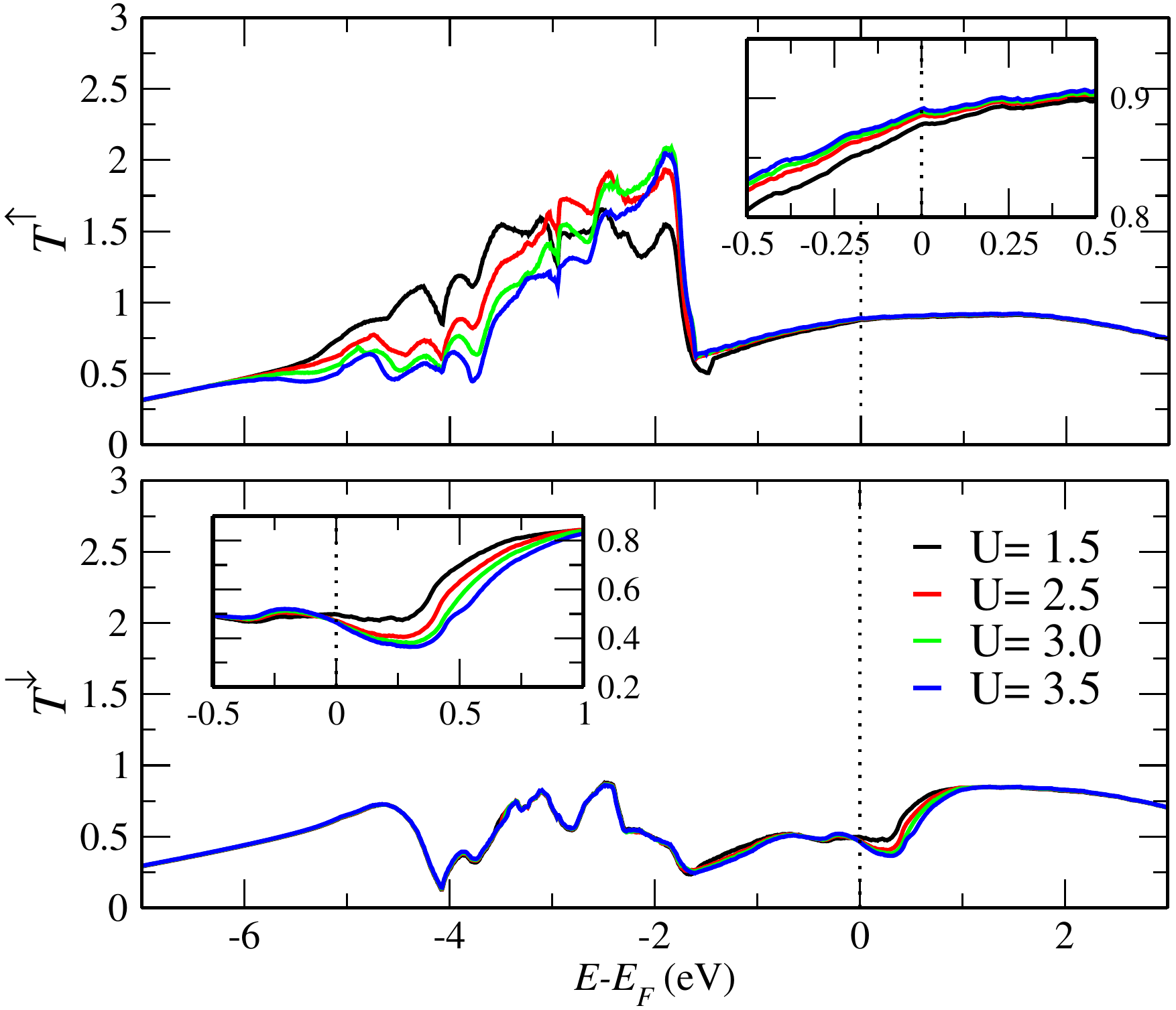}
\caption{Spin up and spin down transmission coefficients as a function of the energy for the Cu/Co/Cu system calculated by means of LSDA+$U$ with $U$ equal to $1.5$ eV (black line), $2.5$ eV (red line), $3$ eV (green line) and $3.5$ eV (blue line). $J$ is constant and equal to $0.5$ eV. The insets zoom in the energy region near the Fermi level. }
\label{fig.TRC_LDAU}
\end{figure}

The transmission coefficient is displayed in Fig. \ref{fig.TRC_LDAU} for the same $U$ values used before. In general we find that there is no drastic reduction of the transmission compared to the LSDA case as the static mean-field potential does not account for relaxation times, differently from the case of the dynamical many-body self-energy. The transmission can be easily related to the main features in the DOS. In particular, we focus on the energy region near the Fermi energy (see inset in Fig. \ref{fig.TRC_LDAU}). In the majority channel, the transport is mostly due to the $s$ electrons. Their scattering with the $d$ states is systematically reduced as these states move towards lower energies when increasing $U$. As a results, $T^\uparrow(E_F)$ increases. 
This behaviour is the opposite with respect to that found in DMFT (see Sec. \ref{Sec.Results} and, moreover, Appendix \ref{app.DMFT_U_J}).
In the spin down channel, we found that the shift of the unoccupied states from the Fermi level induced by the static mean-field potential, leads to an increase of the transmission coefficient with $U$. The overall effect is quite small, but it gives a slight increase in $SP(E_F)$, which reaches the largest value of $0.31$ for $U=3.5$ eV.  \\

\begin{figure}[h]
\centering\includegraphics[width=\columnwidth,clip=true]{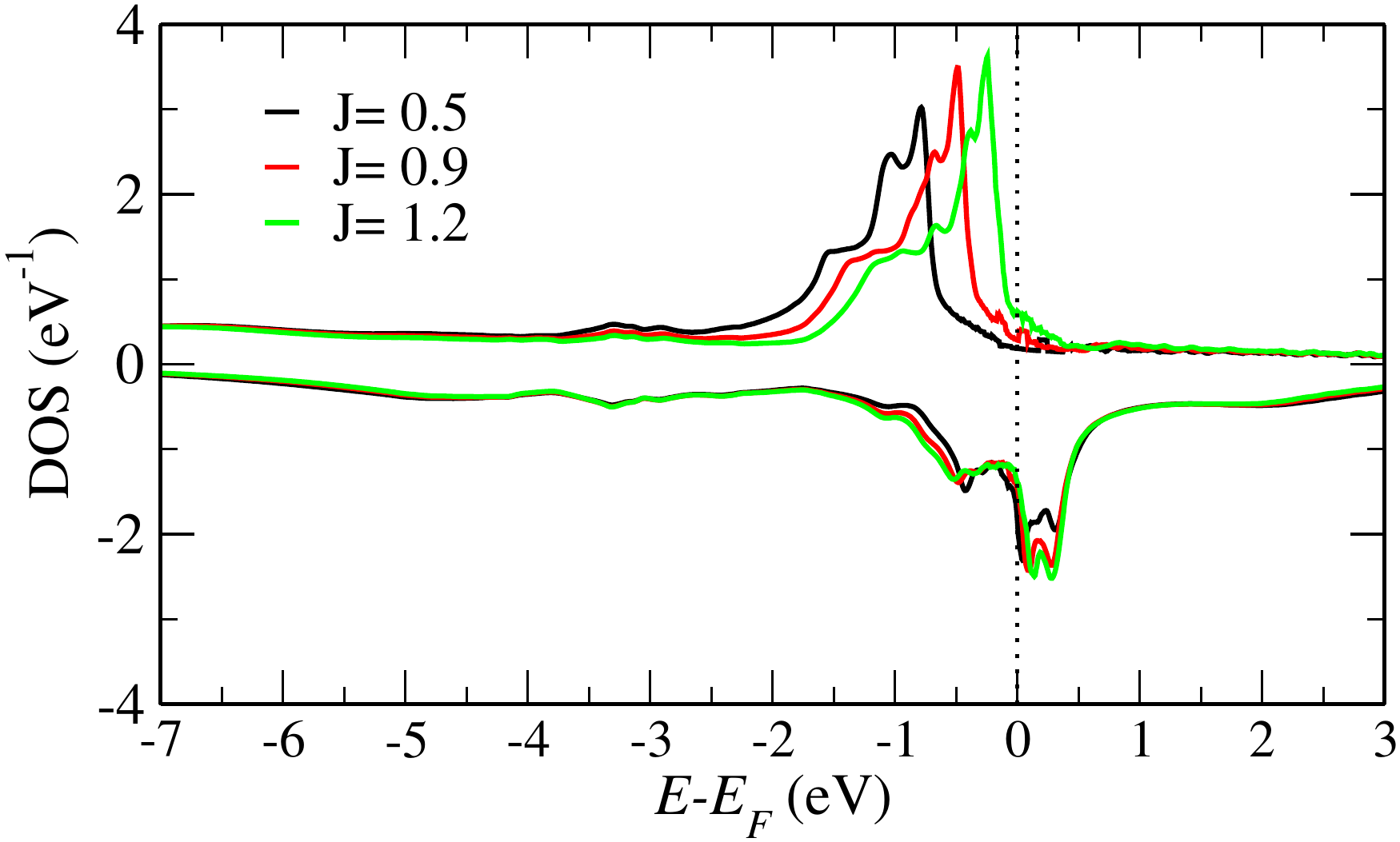}
\caption{DOS of the Co layer in the Cu/Co/Cu system, calculated by using DMFT for $U=3$ eV and $J$ equal to $0.5$ eV (black line), $0.9$ eV (red line), and $1.2$ eV (blue line).}
\label{fig.DOS_DMFT_J}
\end{figure}

\section{$U$- and $J$-dependence of the LSDA+DMFT results}\label{app.DMFT_U_J}
We discuss here the dependence of our LSDA+DMFT results on the strength of the local Coulomb interaction parameters $U$ and $J$. The DOS of the Co atom in Cu/Co/Cu is shown in Fig. \ref{fig.DOS_U_J0.5}-b for $U$ varying from $1.5$ eV to $3.5$ eV, and $J$ fixed at $0.5$ eV. The majority $3d$ states move in energy toward the Fermi level with $U$, while the position of the minority states is much less affected. As a consequence, we find a general reduction of the DOS spin-splitting. Such reduction contrasts the enhancement predicted by the LSDA+$U$ calculations in Appendix \ref{app.LDAU}. Dynamical correlations play a crucial role in counterbalancing static mean-field effects as also discussed in our previous work, Ref. \onlinecite{dr.ra.22}.
Interestingly, although spin down states do not move in energy, some redistribution of their spectral weight occurs resulting in a considerable spectral narrowing in the case of the largest considered $U$ values.\\
A reduction of the DOS spin-splitting is also found when increasing $J$. Fig. \ref{fig.DOS_DMFT_J} displays the results of calculations for $J=0.5$ eV, $J=0.9$ eV and $J=1.2$ eV, and $U$ fixed at $3$ eV. The spin up $3d$ states move toward the Fermi level with $J$, while the spin down DOS is hardly modified. Differently from what we find when varying the $U$ parameter, we note that $J$ has only a very minor effect of the spectral width.\\

\begin{figure}[h]
\centering\includegraphics[width=\columnwidth,clip=true]{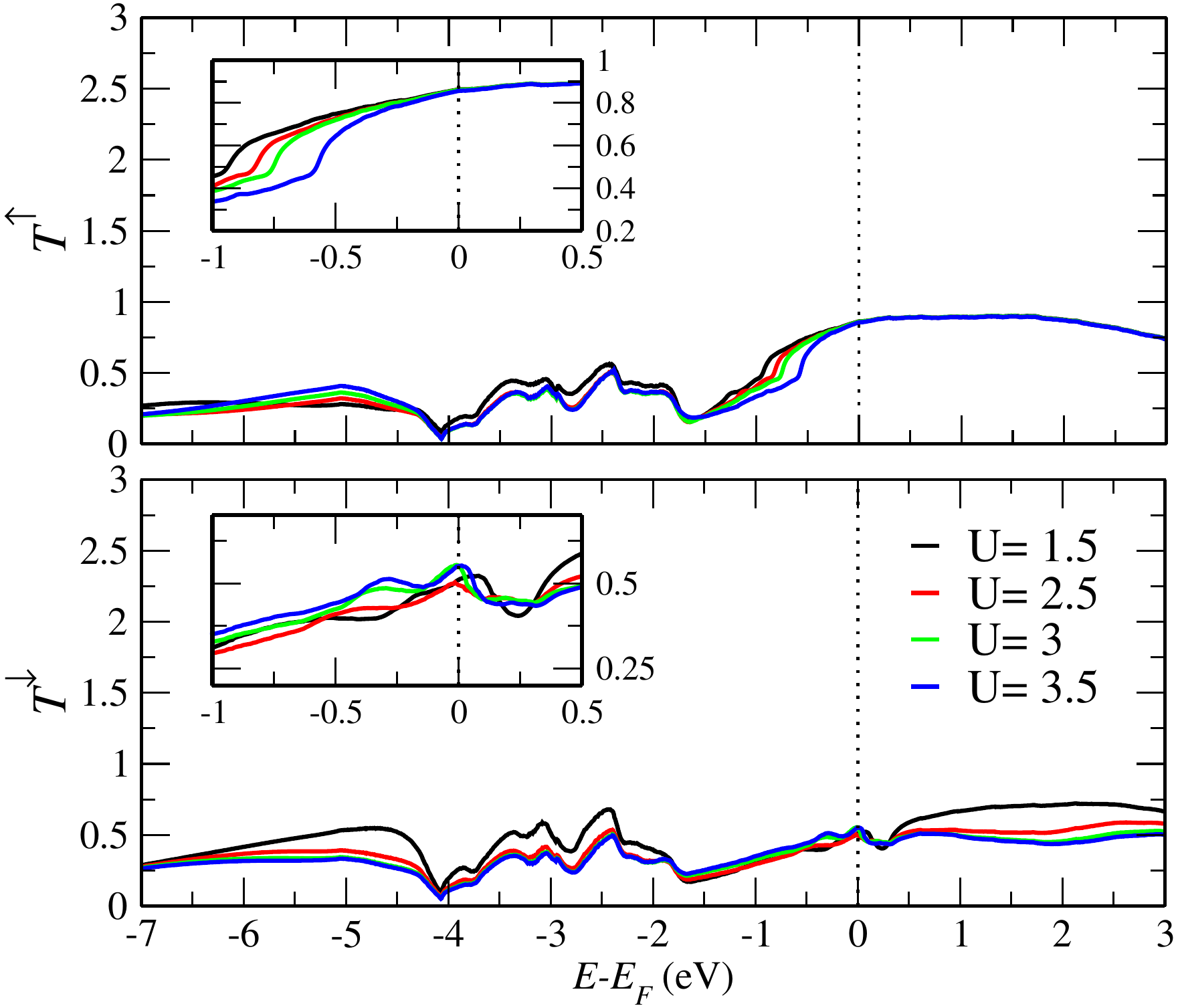}
\caption{Spin up and spin down transmission coefficient for the Cu/Co/Cu system calculated by means of LSDA+DMFT 
with $U$ equal to $1.5$ eV (black line), $2.5$ eV (red line), $3$ eV (green line) and $3.5$ eV (blue line). $J$ is constant and equal to $0.5$ eV. The insets zoom in the energy region near the Fermi level.}
\label{fig.TRC_DMFT_U}
\end{figure}

\begin{figure}[h]
\centering\includegraphics[width=\columnwidth,clip=true]{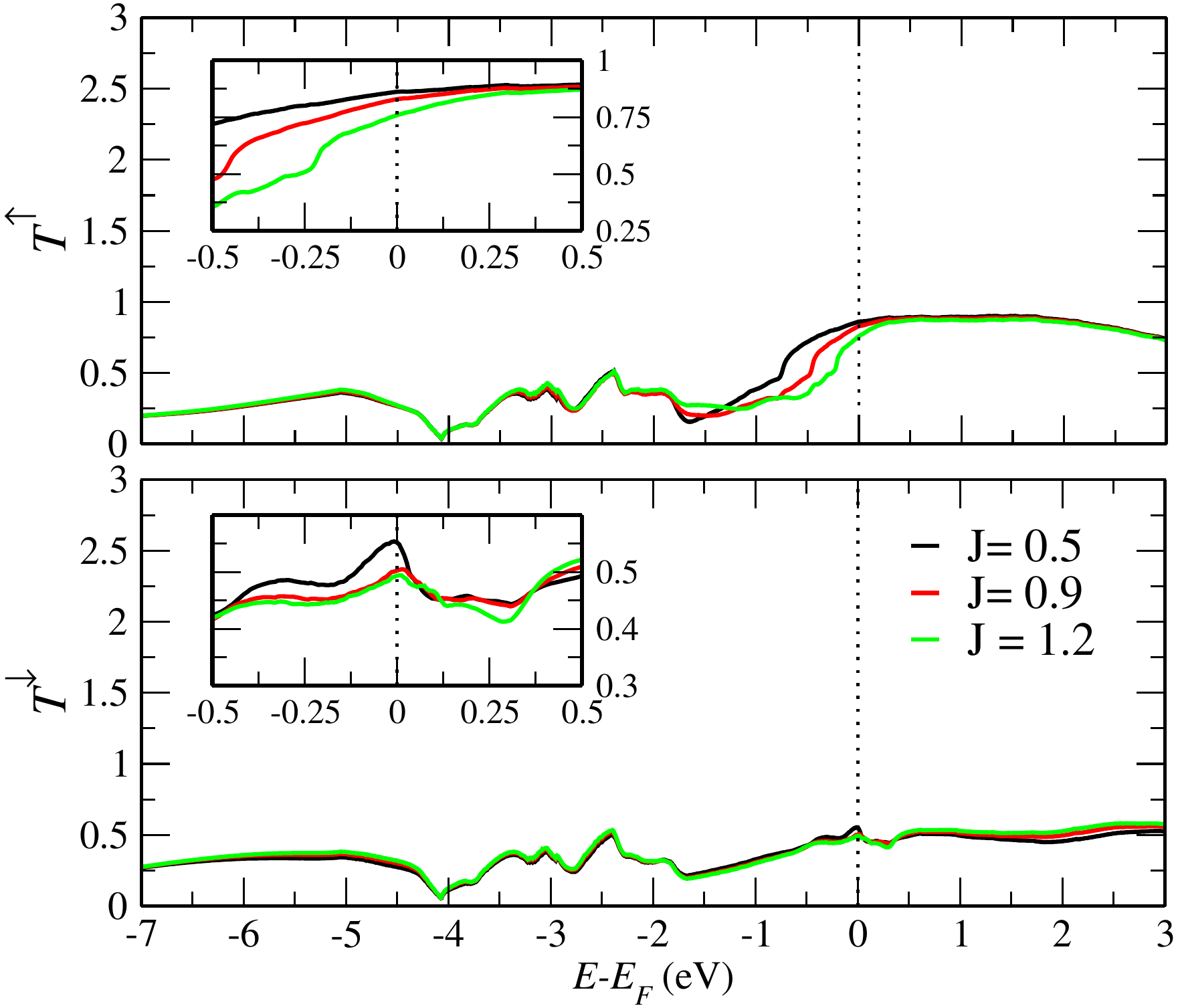}
\caption{Spin up and spin down transmission coefficient for the Cu/Co/Cu system calculated by means of LSDA+DMFT with
$U=3$ eV and $J$ equal to $0.5$ eV (black line), $0.9$ eV (red line), and $1.2$ eV (blue line).
The insets zoom in the energy region near the Fermi level.}
\label{fig.TRC_DMFT_J}
\end{figure}

The spin-dependent DMFT transmission coefficient for Cu/Co/Cu is shown in Fig. \ref{fig.TRC_DMFT_U} for different $U$ values and $J=0.5$ eV. $T^\uparrow(E)$ and $T^\downarrow(E)$ are increasingly suppressed with $U$ in those energy regions, where the transport is through the $d$ states and electronic relaxation is large. This effect is evident in particular for $-4\lesssim E-E_F\lesssim 1.5$ eV in the majority spin channel, and for $E-E_F\lesssim -1.5$ eV and $ E-E_F\gtrsim 0.5$ in the minority channel.
In the other energy regions, the physics is dictated by the scattering of the $s$ electrons with the $3d$ states. The systematic shift in energy of the spin up $3d$ states toward the Fermi level lowers the spin up transmission in the region $-1 \lesssim E-E_F\lesssim -0.2$ eV (see the inset in Fig. \ref{fig.TRC_DMFT_U}). Notably, a similar behavior is also found when increasing $J$ instead of $U$. This is shown in Fig. \ref{fig.TRC_DMFT_J} for $J=0.5$ eV, $J=0.9$ eV, and $J=1.2$ eV and a for $U=3$ eV. The transmission $T^\uparrow(E)$ calculated for these parameters differs mostly at $E-E_F\approx -0.4$ eV. 
At the Fermi energy, any correlation effects are however small. When we compute the spin-polarization defined in Eq. (\ref{eq.SP}), we find a negligible dependence on the interaction strength parameters. The results for $SP(E_F)$ are shown in Table \ref{Tab.SP}. $SP(E_F)$ is very close to the LSDA DFT value, $0.24$, for all considered $U$ and $J$ parameters.     

\begin{table}
{
\begin{tabular}{ M{1.3cm} | M{1.3cm}|M{1.5cm} }\hline
 $U$  &\space  $J$   & \space $SP(E_F)$ \\\hline
 1.5  &  0.5   & 0.26 \\
 2.5  &  0.5   & 0.27  \\
 3.0  &  0.5   & 0.22  \\
 3.5  &  0.5   & 0.21  \\
 3.0  &  0.9   & 0.25  \\
 3.0  &  1.2   & 0.21  \\\hline
  \end{tabular}}
\caption{$SP(E_F)$ calculated by means of DMFT for various $U$ and $J$ values.}\label{Tab.SP}
\end{table}
\begin{figure}[h]
\centering\includegraphics[width=\columnwidth,clip=true]{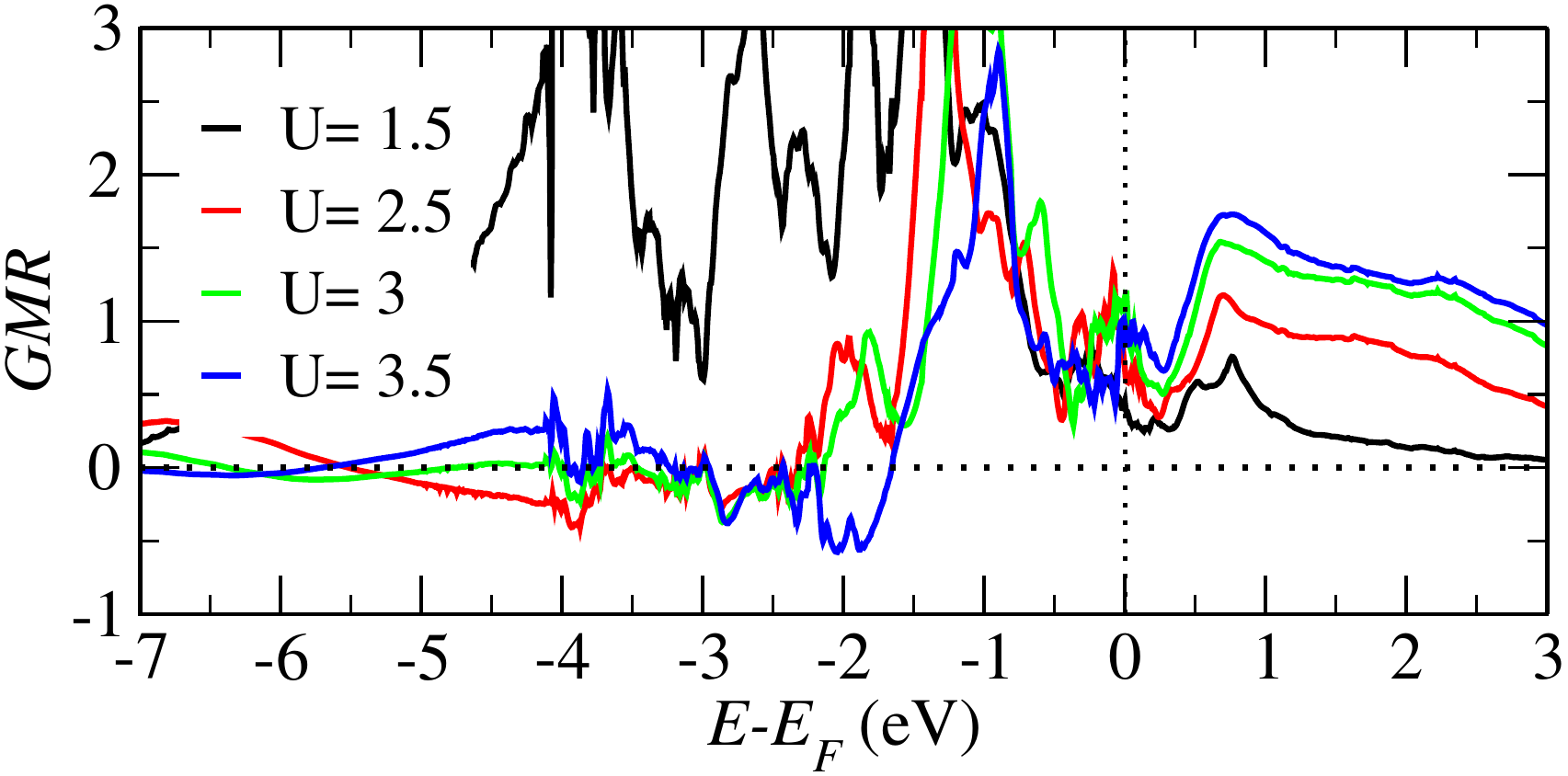}
\caption{$GMR(E)$ calculated by DMFT with $U$ equal to $1.5$ eV (black line), $2.5$ eV (red line), $3$ eV (green line) and $3.5$ eV (blue line).}
\label{fig.GMR_U}
\end{figure}

\begin{table}
{
\begin{tabular}{ M{1.3cm} | M{1.3cm}|M{1.7cm} }\hline
 $U$  &\space  $J$   & \space $GMR(E_F)$ \\\hline
 1.5  &  0.5   & 0.45 \\
 2.5  &  0.5   & 0.68  \\
 3.0  &  0.5   &  1.1 \\
 3.5  &  0.5   &  0.91 \\
 3.0  &  0.9   & 0.45  \\
 3.0  &  1.2   & 0.5  \\\hline
  \end{tabular}}
\caption{$GMR(E_F)$ calculated by means of DMFT for various $U$ and $J$ values.}\label{Tab.GMR_U}
\end{table}
Finally, we assess the calculation of $GMR(E)$ for Cu/Co$_3$/Cu$_3$/Co$_3$/Cu. The results are shown in Fig. \ref{fig.GMR_U} for $U=1.5$ eV, $U=2.5$ eV, $U=3$ eV, $U=3.5$ eV, and $J=0.5$ eV. In the case of small $U$, the shape of the $GMR(E)$ function resembles that computed with DFT in Fig. \ref{fig.GMR}. $GMR(E)$ is large for negative energies, while it drops above the Fermi level. In contrast, when increasing $U$, $GMR(E)$ is drastically suppressed for $E-E_F\lesssim -0.5$ eV, while it systematically increases for positive energies. The maximum at $E-E_F\approx0.7$ eV rises from $0.7$ for $U=1.5$ eV to $1.75$ for $U=3.5$ eV indicating how this feature is related to electron correlations (see also Sec. \ref{Sec.spin_valve}).\\
Spin transport at the Fermi level is strongly dependent on $U$ and also $J$. The values for $GMR(E_F)$ calculated for different parameters are listed in Table \ref{Tab.GMR_U}. They vary from $0.45$ (for $U=1.5$ eV and $J=0.5$ eV) to $1.1$ (for $U=3$ eV and $J=0.5$ eV), which is almost four times the LSDA DFT value, $0.35$. This is because a small energy shift of $3d$ states has a large effects on the transmission coefficient. Therefore, quantitative accurate predictions at the Fermi energy remain rather challenging, and we suggest that studies of linear-response spin transport properties based on LSDA+DMFT should always be accompanied by a careful inspection of the dependence of the results on $U$ and $J$.

\begin{figure}[h]
\centering\includegraphics[width=\columnwidth,clip=true]{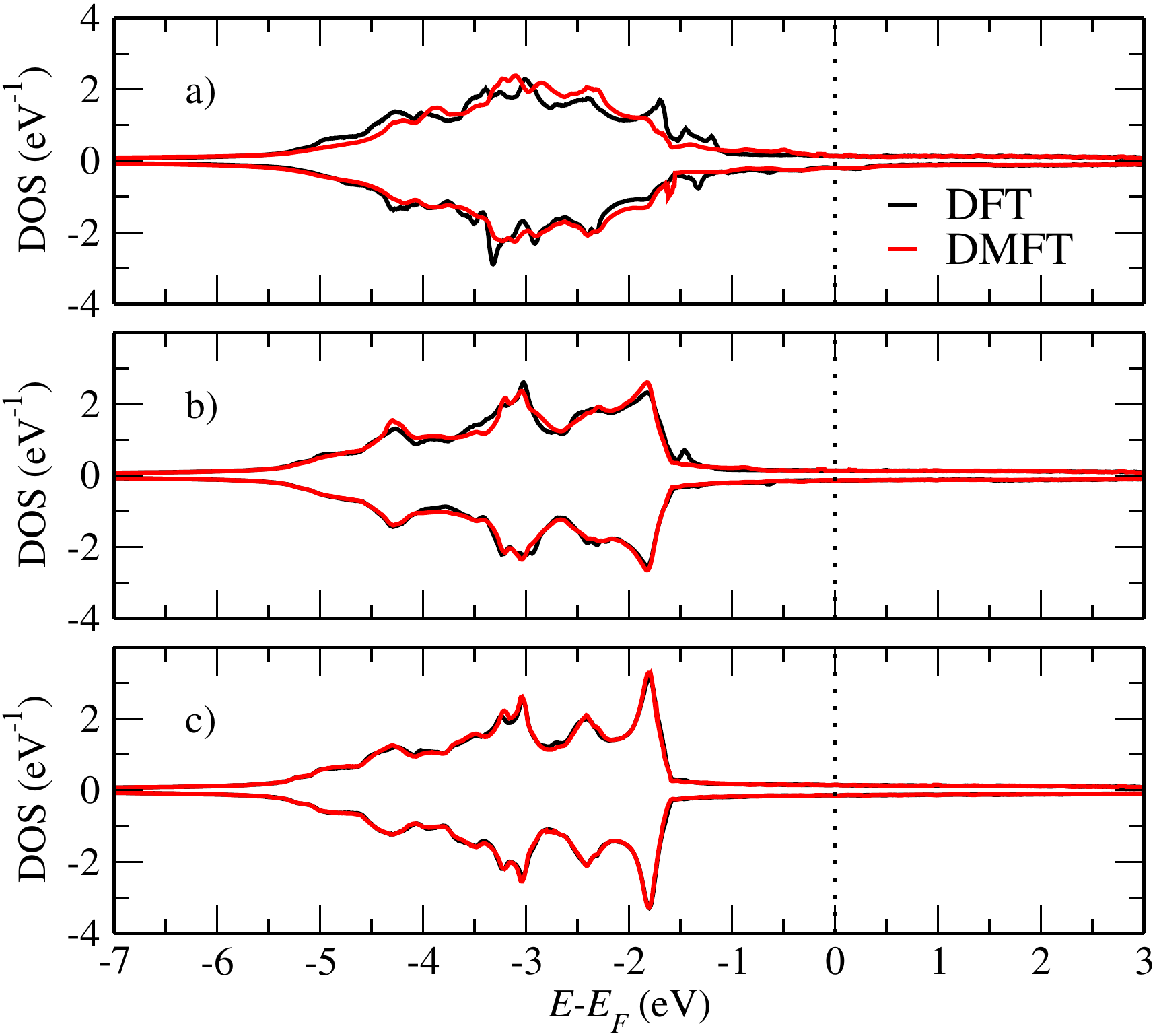}
\caption{DOS of three Cu atoms in the vicinity of the Co layer in Cu/Co/Cu. The Cu atoms are a) nearest neighbor to Co, b) next nearest neighbor to Co, and c) Cu third nearest neighbor to Co. The result of both DFT and DMFT are presented for comparison.}
\label{fig.DOS_Cu}
\end{figure}

\section{DOS of the Cu atoms}\label{app.DOS_Cu}

In the studied
heterostructures, the DOS of the Cu layers in proximity to Co is modified, and a small spin-polarization is induced on both the Cu $s$ and $3d$ states through their hybridization with the Co $3d$ orbitals. We see this effect in Fig. \ref{fig.DOS_Cu} for Cu/Co/Cu. The modification is significant for a Cu atom, which is nearest neighbor to Co (Fig. \ref{fig.DOS_Cu}-a). It becomes small for Cu atoms in the next nearest neighbor position (\ref{fig.DOS_Cu}-b). It is then already negligible for third nearest neighbor Cu atoms (\ref{fig.DOS_Cu}-c), whose DOS closely resembles that of bulk Cu. \\
Even though no on-site interaction terms are added to the Cu atoms, their states are affected by the dynamical self-energy, which ``propogates'' from the Co layer (i.e., the correlated subspace) to the surrounding (i.e., the bath) via the hybridisation and the transformation in Eq. (\ref{SigmaTrans}). The DOS for Cu atoms in the nearest neighbour position to the Co layer present some clear differences in DMFT and DFT calculations (Fig. \ref{fig.DOS_Cu}-a). In particular, the proximity-induced spin-polarization is reduced in DMFT as we can clearly see in the energy region between $E-E_F\approx -2$ eV and -1 eV. Such reduction is a consequence of the smaller Co $3d$ spin-splitting given by DMFT compared to DFT. Correlation effects disappear as the DOS of a Cu atom becomes more bulk-like (\ref{fig.DOS_Cu}-c).    

 \bibliography{ref}

\end{document}